# Reconfigurable multi-tone mmWave/THz generation powered by high-efficiency Kerr parametric oscillation

Dong IL Lee[1,†,*], Fanqi Shen[2,†], Wenzheng Liu[1], Xinghe Jiang[2], Anika Tabassum[2], Alwaleed Aldhafeeri[1], Manas Choure[1], Jasmine Lu[1], Mona Jarrahi[2,*], and Chee Wei Wong[1,*]

[1]Fang Lu Mesoscopic Optics and Quantum Electronics Laboratory, University of California, Los Angeles, CA 90095, USA

[2]Terahertz Electronics Laboratory, University of California, Los Angeles, CA 90095, USA.

[†] These authors contributed equally to this work

[*]Correspondence to: dilee@ucla.edu, mjarrahi@ucla.edu, cheewei.wong@ucla.edu

**Broadband and simultaneously accessible multi-tone millimeter-wave (mmWave) and terahertz (THz) signals are attractive for multi-band wireless communication and spectroscopy applications, yet their generation remains constrained by electronic bandwidth limitations and inefficient optical frequency conversion. Here, we demonstrate reconfigurable photonic generation of multi-tone mmWave/THz signals with comb spacings of 100, 200, 300, and 400 GHz, using a strongly asymmetric dual-ring silicon-nitride (SiN) microresonator and a broadband spiral plasmonic photomixer. A localized avoided-mode crossing (AMX) introduced by an auxiliary ring supports a fixed optical parametric oscillation (OPO) mode pair over an extended pump-detuning range, consistent with passive compensation of detuning-dependent phase mismatch. This enables a maximum pump-to-generated-line conversion efficiency of 67.3 % with 25.06 dB of pump depletion across different pump wavelength selections. Direct photomixing of these optical states produces**

**simultaneous multi-frequency tones spanning 100 to 900 GHz without intermediate optical amplification. The generated carriers are further characterized through down-converted spectra, frequency-noise power spectral density (PSD), and Allan deviation measurements, yielding a β-separation linewidth of 26 kHz at an observation time of 0.1 s. Optical modulation offers high radiated power levels exceeding 2 mW at 100 GHz at a duty cycle of 1 %, while mitigating thermal breakdown in the photomixer. High optical conversion efficiency, reconfigurable and selectable comb spacings, and broadband multi-tone generation on a chip-scale platform provide a compact route toward reconfigurable multi-band mmWave/THz sources for wireless communication, spectroscopy, and sensing.**

## Introduction

The growing demand for high-bandwidth data connectivity is driving the need not only for carrier frequencies extending beyond the W-band [1–3], but also for simultaneous access to multiple widely separated spectral channels. Such multi-frequency operation can increase the aggregate bandwidth of wireless systems through parallel transmission [4–6], while enabling simultaneous interrogation of multiple spectral features in spectroscopy and sensing [7–9]. Access to multiple carriers from a single source can further simplify multi-band operation compared with architectures relying on independently generated frequencies. Optical frequency combs offer a natural route to this functionality because photomixing maps the frequency differences among their equally spaced optical modes onto electrical carriers [10–14]. Microresonator platforms are particularly attractive in this context because their large mode spacings naturally fall within the millimeter-wave (mmWave) and terahertz (THz) bands, and nonlinear optical generation can be realized in compact, on-chip photonic devices [15].

Such compact microresonator sources have enabled various forms of mmWave/THz generation [16–20]. Near-quantum-limited phase noise at frequencies up to 384 GHz [21] and high-power coherent THz radiation with an optical-to-THz conversion efficiency of 1.1 % [22] have been demonstrated using nonlinear optical devices, while broader frequency tunability from 330 GHz to 2.3 THz has been reported using thermally sensitive avoided-mode crossings (AMXs) [23]. More recent approaches have demonstrated a 50-GHz sub-THz baseband comb [24] and dual-soliton-based THz comb synthesis extending beyond 1 THz [25]. Many of these approaches utilize uni-traveling-carrier (UTC) photodiodes, often limited in bandwidth and relying on selected optical line pairs to generate targeted carrier frequencies. On the other hand, plasmonically enhanced photoconductive photomixers can directly convert multiple optical beat notes into free-space radiation over a wide spectral range [26–29]. This broadband conversion is well suited for simultaneous multi-tone generation, allowing multiple optical line pairs to contribute to distinct mmWave/THz frequencies without individual line-pair selection [30]. Nevertheless, simultaneous multi-tone generation with large, reconfigurable frequency spacings of hundreds of gigahertz remains relatively unexplored, particularly when driven directly by a single nonlinear optical source.

Meeting this requirement demands an optical source capable of efficiently distributing substantial optical power among widely separated comb lines. Kerr OPO provides a direct route to achieving this through cavity-enhanced four-wave mixing, which transfers power from a continuous-wave pump to widely separated signal and idler modes [31–33]. Efficient conversion, however, requires favorable phase matching, low cavity loss, suppression of competing nonlinear processes, and efficient output coupling [34,35]. These requirements become particularly stringent in normal-dispersion resonators, where the intrinsic cavity dispersion generally inhibits widely separated OPO states [22,36,37]. Efficient large-spacing oscillation therefore requires targeted resonance shifts to compensate for dispersion-induced phase mismatch.

Coupled-resonator photonic molecules provide such localized control by selectively hybridizing targeted cavity modes through AMX [37–40]. This approach has been used to engineer parametric oscillation and enhance Kerr-comb power-conversion efficiency through selective shifts of cavity resonances [41,42]. During OPO operation, however, the phase-matching condition evolves with pump detuning and intracavity power, as Kerr and thermal shifts modify the hot-cavity mode frequencies [43,44]. Consequently, a mode pair that is phase matched at one operating point can deviate from the optimal phase-matching condition as the nonlinear state evolves. Maintaining a targeted, widely separated OPO mode pair near phase matching over an extended pump-detuning range can therefore sustain high conversion efficiency while broadening the accessible operating window.

Here, we employ a strongly asymmetric dual-ring SiN microresonator in which the auxiliary ring introduces a localized AMX near the targeted OPO mode. As the pump detuning and intracavity power vary, the hybridized resonance shifts to compensate for the evolving hot-cavity phase mismatch, maintaining the same signal–idler mode pair near phase matching over an extended pump-detuning range. Within this operating range, the pump-to-generated-line conversion efficiency reaches 67.3 %, with a pump depletion of up to 25.06 dB. We then use the resulting large-spacing OPO states as the optical input to a broadband plasmonic photomixer. By selecting different pump wavelengths, we achieve four discrete comb spacings of 100, 200, 300, and 400 GHz, generating simultaneous tones spanning 100–900 GHz without intermediate optical amplification. We further characterize the radiated power and frequency stability of the generated carriers across the mmWave and THz bands. These results combine detuning-robust parametric oscillation with broadband multi-tone photomixing, providing a compact approach to reconfigurable multi-band mmWave/THz generation.

## Results

### Concept of a photonic multi-tone mmWave/THz source

Figure 1 illustrates the concept of the multi-tone mmWave/THz source investigated in this work. A continuous-wave optical pump is coupled into a dual-ring SiN microresonator, where Kerr OPO generates multiple optical lines with large frequency separations. The generated optical spectrum is subsequently coupled to a plasmonic photomixer, which converts the frequency differences between the optical lines into multiple mmWave/THz carriers (Fig. 1a). In this architecture, a single nonlinear optical source enables simultaneous access to multiple high-frequency carriers through direct photomixing.

The dual-ring resonator consists of a main ring coupled to a much smaller auxiliary ring (Fig. 1a). The main and auxiliary rings have free spectral ranges (FSRs) of 100.53 GHz and approximately 1 THz, respectively, while the main ring exhibits normal dispersion with $D_2/2\pi$ = -7.86 MHz and an intrinsic quality factor of $Q_{in} = 2.85\times10^6$ for the fundamental transverse-electric ($TE_{00}$) mode. The main ring provides the underlying resonance grid, whereas coupling to the auxiliary-ring mode locally perturbs the resonance spectrum near the targeted OPO wavelength of 1566.38 nm. This mode interaction is evident in the measured integrated dispersion, $D_{int}/2\pi$, where the resonance frequencies exhibit localized deviations from the otherwise smooth normal-dispersion background (Fig. 1a(i)). For a signal–idler pair satisfying the azimuthal mode-number condition, $2m_p = m_s + m_i$, the parametric process further depends on the alignment of the corresponding cavity resonances with the energy-conserving frequencies. We characterize this alignment using the resonance-frequency mismatch, $\Delta\omega_{res} = \omega_{s,res} + \omega_{i,res} - 2\omega_p$, where $\omega_p$ is the pump-laser frequency, and $\omega_{s,res}$, and $\omega_{i,res}$ are the corresponding hot-cavity resonance frequencies. In Fig. 1a(i), $\delta_{rel}$, schematically denotes the frequency splitting between the two hybridized resonances near the targeted OPO wavelength. As the pump detuning changes, redistribution of the intracavity field modifies the Kerr and thermal shifts in the two rings, changing their relative resonance positions. This AMX-mediated resonance response partially compensates for the detuning-dependent variation in $\Delta\omega_{res}$, maintaining favorable resonance alignment for the targeted signal–idler pair over an extended pump-detuning range. The resulting large-spacing OPO spectrum is directly converted into THz radiation by a broadband plasmonic photomixer fabricated on an ErAs:InGaAs substrate (Fig. 1a(ii)). Plasmonic contact electrodes at the antenna input reduce the photocarrier transport distance and enhance the optical-to-mmWave/THz conversion efficiency by converting beat notes between pairs of optical lines at frequencies $\nu_m$ and $\nu_n$ into radiation at $\nu_{mmWave/THz} = \Delta\nu_{mn} = |\nu_m - \nu_n|$. A multi-line OPO spectrum therefore simultaneously generates multiple mmWave/THz carriers corresponding to all pairwise frequency differences.

Figure 1b extends this concept to a multi-band mmWave/THz transmitter architecture. The chip-scale photonic source provides multiple widely separated optical carriers that can be independently modulated prior to photomixing. Parallel photomixing then converts the optical channels into distinct mmWave/THz bands, enabling simultaneous multi-frequency transmission from a common photonic oscillator. This conceptual architecture illustrates how the multi-tone source demonstrated in this work could be integrated into a photonic transmitter for reconfigurable multi-band operation.

## Reconfigurable multi-tone mmWave/THz generation

To generate and characterize the reconfigurable multi-tone mmWave/THz signals, the optical output from the dual-ring OPO is directly coupled to a plasmonic photomixer, as illustrated in Fig. 2a. A free-running continuous-wave pump from an external-cavity diode laser (ECDL) is amplified by an erbium-doped fiber amplifier (EDFA), polarization-controlled to excite the $TE_{00}$ mode, and coupled into the dual-ring resonator. The on-chip pump power is approximately 23 dBm, accounting for a single-facet coupling loss of approximately 2 dB. A portion of the resonator output is monitored using an optical spectrum analyzer (OSA), while the remaining optical power is directed to the photomixer without intermediate optical amplification. The emitted mmWave/THz radiation is collimated and collected using a pair of parabolic mirrors and subsequently down-converted using harmonic mixers (Eravant and VDI) covering the respective frequency bands. The down-converted intermediate-frequency (IF) signal is referenced to an oven-controlled crystal oscillator (OCXO; R&S FSWP26) and analyzed using an RF spectrum analyzer (E4448A, Agilent). Representative microscope images of the dual-ring resonator and plasmonic photomixer are shown in Fig. 2d.

The frequency spacing of the generated spectrum is reconfigured by tuning the pump wavelength to 1565.56, 1564.74, 1563.92, and 1563.10 nm, providing access to OPO states with frequency spacings of approximately 100, 200, 300, and 400 GHz, respectively (Fig. 2b). In each state, the primary OPO pair is accompanied by additional optical lines lying on the same frequency grid, providing multiple optical frequency differences within a single spectrum. Direct photomixing of the approximately 40 mW optical output consequently maps this optical frequency grid onto the mmWave/THz domain. For example, the 300-GHz-spaced OPO state generates signals at 300, 600, and 900 GHz, while the 400-GHz-spaced state produces corresponding tones at 400 and 800 GHz (Fig. 2c). As the OPO spacing decreases, more frequency orders become accessible within the same spectral window, with the 100-GHz-spaced state yielding the densest set of measured tones at successive 100-GHz intervals. Notably, the individual spectra in Fig. 2c are acquired sequentially because of the limited operating bandwidth of each harmonic mixer, although all corresponding mmWave/THz tones coexist simultaneously for a given OPO state. The measured mmWave/THz spectra therefore preserve the frequency grid established by the optical OPO states, with the multi-tone spacing reconfigured from 100 to 400 GHz through pump-wavelength selection.

## High-efficiency OPO in the dual-ring resonator

To investigate the nonlinear evolution of the dual-ring OPO, we use the 400-GHz-spaced OPO state as a representative case. The optical spectrum and hot-cavity response are measured as the pump detuning is varied using the experimental configuration shown in Fig. 3a. A counter-propagating probe laser scans the cavity transmission [42] near the targeted OPO wavelength of 1566.38 nm, with the frequency axis calibrated using a Mach–Zehnder interferometer (MZI). The generated optical spectrum is measured using an optical spectrum analyzer (OSA) to track the OPO evolution (Fig. 3b), while the probe transmission and MZI signals are recorded using an oscilloscope to characterize the hot-cavity response (Fig. 3c). These measurements enable direct comparison of the OPO state and cavity response under the same nonlinear operating conditions.

Following the onset of the 400-GHz OPO, optical power is transferred from the pump to the primary OPO pair, followed by the formation of additional lines on the same frequency grid. The primary OPO spacing remains nearly unchanged as the pump detuning increases, indicating that the same signal-idler mode pair is maintained without switching to competing OPO states.

Efficient OPO operation persists over a pump-detuning range of nearly 18 GHz, demonstrating robust operation without transitions to subsequent nonlinear states and sustained favorable phase matching across a broad detuning range. Over the same range, the hot-cavity resonance remains observable in the probe transmission near the OPO wavelength (Fig. 3c). The extracted mode splitting ($\delta_{rel}$) ranges from 168.4 to 190.9 MHz throughout the scan (Supplementary Note 2), while the accompanying microscope images reveal a redistribution of the optical field between the main and auxiliary rings with increasing pump detuning. The persistent mode splitting and sustained OPO state are consistent with the localized mode interaction compensating for the detuning-dependent resonance mismatch.

The conversion efficiencies measured over the same pump-detuning scan are shown in Fig. 3d, following the procedure described in [42], in which the off-resonance transmitted pump power and the generated optical powers are used to estimate the conversion efficiencies. We define the total and primary-pair conversion efficiencies as: $\eta_{tot} = \frac{P_{\mathrm{gen}}}{P_{\mathrm{p,off}}}$ and $\eta_{pri} = \frac{P_{\mathrm{s}}+P_i}{P_{\mathrm{p,off}}}$, where $P_{\mathrm{p,off}}$ is the transmitted pump power measured with the cavity off resonance, $P_{gen}$ is the total power in all generated lines excluding the residual pump, and $P_s$ and $P_i$ are the powers of the primary signal and idler, respectively. At a pump detuning of -8.7 GHz, marked as operating point (i), the total conversion efficiency reaches 67.3 %, with a primary-pair conversion efficiency of 58.3 %. The total conversion efficiency remains above 60 % over a pump-detuning range of approximately -7.0 to -13.3 GHz, corresponding to a 6.3-GHz operating window. Thus, high-efficiency conversion is sustained over a substantial detuning range rather than being confined to a narrow operating point. The maximum pump depletion occurs at a pump detuning of -11.85 GHz, marked as operating point (ii), where the residual pump is suppressed by 25.06 dB (Supplementary Note 3). The 3.15-GHz separation between the conversion-efficiency and pump-depletion maxima indicates that maximum pump extinction and maximum power transfer to the generated OPO lines occur under different operating conditions. This separation suggests further opportunities to improve conversion efficiency through optimization of the cavity coupling and mode interaction. Nevertheless, the primary-pair efficiency measured at a 400-GHz OPO separation exceeds prior values in the mmWave/THz range and approaches the high efficiencies reported for OPO separations of several terahertz (Fig. 3f) [45–51].

Beyond the representative 400-GHz OPO state, the same dual-ring resonator supports OPO frequency separations ranging from approximately 100 GHz to 1.2 THz. Pumping successive cavity resonances produces these states in approximately 100-GHz increments, as shown in the corresponding optical spectra in Fig. 3e. Across the different pump wavelengths, one of the primary OPO sidebands remains near the AMX region around 1566.38 nm, while the pump and complementary sideband shift according to the selected cavity resonance. This behavior enables access to widely separated optical states that serve as the basis for reconfigurable multi-tone mmWave/THz generation.

**Broadband and continuously tunable mmWave/THz generation**

Next, the frequency coverage and output characteristics of the photonic mmWave/THz source are evaluated using the experimental configurations shown in Fig. 4a. For frequency-domain characterization, selected optical lines ($\nu_{pump}$ and $\nu_{AMX}$) from the OPO spectrum are separated using a wavelength-division multiplexer (WDM), amplified by an EDFA, and coupled to the photomixer (Fig. 4a(i)). To extend the accessible pump-detuning range and thereby increase the continuous frequency-tuning range, the on-chip pump power is increased to approximately 25 dBm, resulting in a broader thermal triangle. The generated mmWave/THz signal is then collected in free space and down-converted using harmonic mixers covering the corresponding frequency bands, following the procedure described previously. This configuration is used to characterize individual frequency components generated by the photomixer, complementing the simultaneous multi-tone generation demonstrated in Fig. 2.

The measured down-converted spectra are shown in Fig. 4b. Nine carrier frequencies spanning 100 GHz to 1.1 THz are experimentally demonstrated, verifying broadband photonic generation over more than one decade in frequency. In addition to the discrete frequency selection enabled by the OPO spacing, each carrier can be continuously tuned by varying the pump detuning. As shown in Fig. 4c, a continuous tuning range exceeding 600 MHz is achieved for carrier frequencies from 100 to 800 GHz, with a maximum tuning span of 850 MHz at 700 GHz. For the 1.0- and 1.1-THz carriers, the tuning ranges are approximately 200 MHz and 480 MHz, respectively. We attribute the absence of the 500- and 900-GHz states to parasitic TE–TM mode coupling near the pump wavelength, which perturbs the cavity resonances and disrupts the corresponding phase-matching conditions. Furthermore, for OPO spacings beyond 1 THz, the achievable continuous tuning range becomes narrower due to periodic hybridization between the main- and auxiliary-ring modes, associated with their approximately 1:10 FSR ratio. This mode interaction reduces the stability of the OPO state against variations in pump detuning, limiting the detuning range over which the same signal–idler mode pair can be maintained.

The radiated output power is enhanced and characterized using the experimental configuration shown in Fig. 4a(ii). To increase the optical drive while limiting the average thermal load on the photomixer, the optical excitation is modulated using an acousto-optic modulator (AOM) operating at 250 kHz with a duty cycle of 1 % [27]. Under this modulated operating condition, an average optical power of approximately 70 mW is coupled into the photomixer. The emitted THz radiation is collected and measured using a calibrated pyroelectric detector (Sensors und Lasertechnik, THz-30). During the on-state of each modulation cycle, the photomixer generates THz output powers of 2.086 mW at 100 GHz and 150 μW at 1.1 THz (Fig. 4d), demonstrating broadband photomixing across the mmWave/THz frequency range.

**Frequency-noise and stability characterization of generated mmWave/THz carriers**

The spectral characteristics of the generated mmWave/THz carriers are further evaluated through frequency-noise measurements. Figure 5a shows the frequency-noise power spectral densities (PSDs) of representative carriers at 300, 600, and 700 GHz. Despite the substantial differences in carrier frequency, the three signals exhibit similar absolute frequency-noise levels over most of the measured offset-frequency range [52]. This behavior indicates that the frequency fluctuations do not simply scale with the generated carrier frequency. For a photomixed carrier, the instantaneous frequency fluctuation is determined by the difference between the fluctuations of the two optical frequencies, $\delta\nu_{mmWave/THz} = \delta|\nu_1 - \nu_2|$, such that common-mode resonance fluctuations can be partially suppressed [53]. In the present dual-ring OPO, one primary sideband remains localized near the same AMX region across the different OPO states with frequency spacings ranging from 100 to 800 GHz. The similar noise spectra are therefore consistent with a common cavity-mediated fluctuation mechanism, potentially associated with the thermo-optic response of the hybridized mode and the accompanying redistribution of the intracavity field [43,44,54]. Using the $\beta$-separation method [55], the measured frequency-noise PSDs yield linewidths of 26.1, 41.1, and 33.1 kHz for the 300-, 600-, and 700-GHz carriers, respectively, when integrated down to an offset frequency of 10 Hz. At higher offset frequencies, the measured noise approaches the detection floor of the harmonic-mixer measurement chain. The extracted linewidths are therefore reported as measured system-level values rather than intrinsic linewidth limits.

The longer-term frequency stability is characterized using an 800-GHz carrier, as shown in Fig. 5b. For short averaging times, the Allan deviation is calculated from the measured frequency-noise PSD, whereas direct frequency-counter measurements are used for longer averaging times. The Allan deviation reaches a minimum of $3.8 \times 10^{-9}$ at an averaging time of $3.4 \times 10^{-5}$ s, beyond which it increases with longer averaging times as slower frequency fluctuations become dominant. The continuous frequency trace in Fig. 5c further illustrates the free-running behavior of the 800-GHz carrier over approximately 30 min. During this interval, the frequency fluctuations exhibit a standard deviation of approximately 312 kHz. Together, these measurements characterize both the short-term frequency stability and the longer-term free-running behavior of the generated mmWave/THz carriers. While these results demonstrate exceptional frequency stability for a free-running chip-scale system, suitable for wireless communication and many spectroscopy applications, the stability can be further improved by locking the system to reference cavities [13,53] for precision spectroscopy and metrology applications, albeit at the expense of increased system size, weight, and power (SWaP).

## Discussion

We have demonstrated a reconfigurable multi-tone mmWave/THz source based on a high-efficiency dual-ring SiN Kerr OPO and broadband photomixing. Localized mode interaction near the targeted OPO wavelength enables the same signal–idler pair to remain accessible over nearly 18 GHz of pump detuning, with the pump-to-generated-line conversion efficiency reaching 67.3 % and remaining above 60 % over a 6.3-GHz detuning range. The maximum pump depletion reaches 25.06 dB at a different operating point, indicating further potential for optimizing the transfer of pump power into the generated OPO lines. The high conversion efficiency provides sufficient optical power for direct multi-tone photomixing without intermediate optical amplification, while pump-mode selection enables reconfiguration of the OPO spacing in approximately 100-GHz increments.

The resulting optical states are directly converted into broadband, tunable mmWave/THz signals. Pump-wavelength selection enables multi-tone spacings ranging from 100 to 400 GHz, while direct photomixing of the multi-line OPO spectra generates simultaneous carrier frequencies extending to 900 GHz. Individual carrier generation is further demonstrated over the 100-GHz to 1.1-THz frequency range, along with continuous frequency tuning of approximately 850 MHz within a selected OPO state. The present frequency coverage is primarily limited by the resonator modal structure, including parasitic TE–TM interactions and periodic main–auxiliary-ring hybridization at larger OPO spacings, rather than by the broadband photomixing process itself. Further suppression of these parasitic mode interactions, combined with optimization of the cavity coupling and auxiliary-ring spectrum, could extend the accessible frequency range and improve conversion efficiency.

Another important advantage of the demonstrated system is its high frequency stability without requiring external reference-cavity stabilization. This free-running stability is particularly attractive for high-data-rate wireless communication, where low phase noise and stable carriers are essential for coherent transmission and spectrally efficient modulation, as well as for high-resolution spectroscopy and sensing applications. Eliminating bulky reference cavities and associated stabilization electronics significantly reduces system size, weight, power, and complexity, offering a promising pathway toward chip-scale, low-SWaP mmWave/THz systems.

These results establish a pathway toward chip-scale, multi-band mmWave/THz sources based on high-efficiency, reconfigurable Kerr OPO generation. The OPO can be heterogeneously integrated with photomixers, pump lasers, semiconductor optical amplifiers, and modulators [56], providing a scalable platform for future high-data-rate communication and spectroscopy systems.

## Methods

### Dual-ring resonator design and characterization

The device is based on a 350-nm-thick SiN platform and consists of a main ring coupled to a smaller auxiliary ring. The main ring has a radius of 238.15 μm and a waveguide width of 2.0 μm, while the auxiliary ring has a radius of 23.79 μm and a waveguide width of 3.5 μm, corresponding to free spectral ranges of 100.53 GHz and approximately 1 THz, respectively. The ring-to-ring coupling gap is designed to be 350 nm, while the fundamental $TE_{00}$ mode of the main ring exhibits normal dispersion, with $D_2/2\pi = -7.86$ MHz and an intrinsic quality factor of $2.85\times10^6$. The resonance frequencies were measured using a hydrogen-cyanide-referenced swept-wavelength interferometer (SWI) [22,23] and used to calculate the integrated dispersion according to $D_{int}(\mu)=\omega_\mu-(\omega_0+D_1\mu)$, where $\mu$ is the integer mode index relative to the reference resonance $\omega_0$, $\omega_\mu$ is the angular frequency of mode $\mu$, and $D_1/2\pi$ is the main-ring FSR. Coupling to the auxiliary-ring mode produces a localized avoided-mode crossing near 1566.38 nm, perturbing the otherwise smooth normal-dispersion resonance spectrum and defining the targeted OPO operating region used throughout the measurements.

### Photomixer fabrication

The plasmonic contact electrodes are defined by electron-beam lithography, followed by the deposition of 3/97 nm Ti/Au and subsequent liftoff. A 240-nm-thick $Si_3N_4$ antireflection coating is then deposited using plasma-enhanced chemical vapor deposition (PECVD). Subsequently, contact vias are defined by optical lithography and opened by reactive-ion etching of the $Si_3N_4$ layer. Finally, the logarithmic spiral antenna and bias lines are patterned using optical lithography, followed by the deposition of 50/450 nm Ti/Au and liftoff. Optical microscopy and scanning electron microscopy (SEM) images of the fabricated device are shown in Fig. 2d.

**OPO and hot-cavity characterization**

Optical parametric oscillation is generated using a tunable ECDL (Newport TLB6700), amplified by an EDFA and polarization-controlled to excite the $TE_{00}$ mode of the resonator. The nominal on-chip pump power is 23 dBm, accounting for a single-facet coupling loss of approximately 2 dB, while an on-chip pump power of approximately 25 dBm is used for the extended-detuning measurements in Fig. 4. For hot-cavity characterization, a second probe laser is injected counter-propagating to the pump and scanned across the cavity resonances near the targeted OPO wavelength of 1566.38 nm. The probe-frequency axis is calibrated using an MZI with a fringe spacing of 3.95 MHz, and the probe transmission and MZI signals are simultaneously recorded using an oscilloscope, while the OPO spectrum is measured using an OSA at the corresponding pump-detuning condition. The local mode splitting, $\delta_{rel}$, is determined from the frequency separation between the two transmission minima, with the extraction procedure detailed in Supplementary Note 2. For the conversion-efficiency measurements, the off-resonance transmitted pump power, measured under the same optical coupling conditions, is used as the input-power reference, following the procedure described in the Results section.

**Generated mmWave/THz characterization**

For simultaneous multi-tone measurements, the complete OPO spectrum, with a total optical power of approximately 40 mW, was coupled directly to the broadband photomixer without post-resonator optical amplification. The radiated signal was collected using free-space optics (Thorlabs MPD239-M01) and a horn antenna (Eravant SAR-series). Harmonic mixers from Eravant and Virginia Diodes Inc. (VDI), selected according to their respective operating frequency bands, were used to down-convert the generated mmWave/THz carriers to intermediate frequencies (IFs). The resulting IF signals were amplified using a 50-dB RF amplifier (RF Bay LNA1050), filtered using a 10-MHz RF bandpass filter (BPF), and analyzed using an RF spectrum analyzer (RFSA; Agilent E4448A) with a resolution bandwidth (RBW) of 2.4 kHz, referenced to a phase-noise analyzer (PNA; Rohde & Schwarz FSWP26). For individual-carrier generation and frequency-tuning measurements, selected optical lines were isolated using an optical BPF, amplified by an EDFA to a total optical power of up to 40 mW, and delivered to the photomixer, following the same measurement chain used for the multi-tone measurements. The OCXO of the PNA was used as the frequency reference for the harmonic mixers. Radiated-power measurements were performed separately using a pyroelectric detector (Sensors und Lasertechnik THz-30). The optical excitation before THz radiation was modulated at 14 Hz using a mechanical chopper, and the pyroelectric detector output was demodulated at the same frequency using a lock-in amplifier (Zurich Instruments) to determine the radiated THz power. For frequency-stability characterization, the phase-noise spectra of the down-converted carriers were measured and converted into frequency-noise PSDs, as presented in Fig. 5a. For short-term stability analysis, the Allan deviation was calculated from the fractional-frequency noise PSD, $S_y(f)$, according to $\sigma_y(\tau) = \left[2\int S_y(f)\frac{sin^4(\pi f\tau)}{(\pi f\tau)^2}df\right]^{1/2}$ [57], where τ is the averaging time. The PSD-derived Allan deviation was used for short averaging times, while direct frequency-counter measurements (Keysight 53220A) were used for longer-term stability characterization.

## References

[1] T. Nagatsuma, G. Ducournau, and C. C. Renaud, Advances in terahertz communications accelerated by photonics, Nat. Photonics **10**, 371 (2016).

[2] K. Liu, Y. Feng, C. Han, B. Chang, Z. Chen, Z. Xu, L. Li, B. Zhang, Y. Wang, and Q. Xu, High-speed 0.22 THz communication system with 84 Gbps for real-time uncompressed 8K video transmission of live events, Nat. Commun. **15**, 8037 (2024).

[3] Y. Tokizane et al., Beyond 350 GHz: Single-channel 112 Gbps photonic wireless transmission at 560 GHz using soliton microcombs, Commun. Eng. **5**, 77 (2026).

[4] H. Shams, M. J. Fice, K. Balakier, C. C. Renaud, F. Van Dijk, and A. J. Seeds, Photonic generation for multichannel THz wireless communication, Opt. Express **22**, 23465 (2014).

[5] X. Yu, S. Jia, H. Hu, M. Galili, T. Morioka, P. U. Jepsen, and L. K. Oxenløwe, 160 Gbit/s photonics wireless transmission in the 300-500 GHz band, APL Photonics **1**, 081301 (2016).

[6] S. Jia, X. Yu, H. Hu, J. Yu, T. Morioka, P. U. Jepsen, and L. K. Oxenløwe, 120 Gb/s Multi-Channel THz Wireless Transmission and THz Receiver Performance Analysis, IEEE Photonics Technol. Lett. **29**, 310 (2017).

[7] T. Yasui, Y. Kabetani, E. Saneyoshi, S. Yokoyama, and T. Araki, Terahertz frequency comb by multifrequency-heterodyning photoconductive detection for high-accuracy, high-resolution terahertz spectroscopy, Appl. Phys. Lett. **88**, 241104 (2006).

[8] I. A. Finneran, J. T. Good, D. B. Holland, P. B. Carroll, M. A. Allodi, and G. A. Blake, Decade-Spanning High-Precision Terahertz Frequency Comb, Phys. Rev. Lett. **114**, 163902 (2015).

[9] X. Ma, Z. Qin, X. Bi, Z. Li, W. Wan, B. Liu, G. Liu, Y. Pan, Y. Lu, and H. Li, Reconfigurable Terahertz Multi-Harmonic Dual-Combs, Nanophotonics **15**, e70281 (2026).

[10] S. A. Diddams, K. Vahala, and T. Udem, Optical frequency combs: Coherently uniting the electromagnetic spectrum, Science **369**, eaay3676 (2020).

[11] A. Aldhafeeri, H.-H. Chin, T. Melton, D. I. Lee, A. Chu, W. Wang, M. Yu, P. G.-Q. Lo, D.-L. Kwong, and C. W. Wong, Low phase noise K-band signal generation using polarization diverse single-soliton integrated microcombs, Photonics Res. **12**, 1175 (2024).

[12] J. Yang, D. Il Lee, D.-C. Shin, J. Lee, B. S. Kim, H. J. Kang, Y.-J. Kim, and S.-W. Kim, Frequency comb-to-comb stabilization over a 1.3-km free-space atmospheric optical link, Light Sci. Appl. **11**, 253 (2022).

[13] D.-C. Shin, B. S. Kim, H. Jang, Y.-J. Kim, and S.-W. Kim, Photonic comb-rooted synthesis of ultra-stable terahertz frequencies, Nat. Commun. **14**, 790 (2023).

[14] G. Kang et al., Nanometre-precision terahertz interferometry for battery electrode metrology, Nat. Commun. **17**, 7382 (2026).

[15] P. Del'Haye, A. Schliesser, O. Arcizet, T. Wilken, R. Holzwarth, and T. J. Kippenberg, Optical frequency comb generation from a monolithic microresonator, Nature **450**, 1214 (2007).

[16] T. Tetsumoto, T. Nagatsuma, M. E. Fermann, G. Navickaite, M. Geiselmann, and A. Rolland, Optically referenced 300 GHz millimetre-wave oscillator, Nat. Photonics **15**, 516 (2021).

[17] B. Wang, J. S. Morgan, K. Sun, M. Jahanbozorgi, Z. Yang, M. Woodson, S. Estrella, A. Beling, and X. Yi, Towards high-power, high-coherence, integrated photonic mmWave platform with microcavity solitons, Light Sci. Appl. **10**, 4 (2021).

[18] N. Kuse, K. Nishimoto, Y. Tokizane, S. Okada, G. Navickaite, M. Geiselmann, K. Minoshima, and T. Yasui, Low phase noise THz generation from a fiber-referenced Kerr microresonator soliton comb, Commun. Phys. **5**, 312 (2022).

[19] S. Sun et al., Integrated optical frequency division for microwave and mmWave generation, Nature **627**, 540 (2024).

[20] S. Sun et al., Microcavity Kerr optical frequency division with integrated SiN photonics, Nat. Photonics **19**, 637 (2025).

[21] K. Jia et al., Low-noise frequency synthesis and terahertz wireless communication driven by compact turnkey Kerr combs, Nat. Commun. **16**, 6253 (2025).

[22] S.-W. Huang, J. Yang, S.-H. Yang, M. Yu, D.-L. Kwong, T. Zelevinsky, M. Jarrahi, and C. W. Wong, Globally Stable Microresonator Turing Pattern Formation for Coherent High-Power THz Radiation On-Chip, Phys. Rev. X **7**, 041002 (2017).

[23] W. Wang, P.-K. Lu, A. K. Vinod, D. Turan, J. F. McMillan, H. Liu, M. Yu, D.-L. Kwong, M. Jarrahi, and C. W. Wong, Coherent terahertz radiation with 2.8-octave tunability through chip-scale photomixed microresonator optical parametric oscillation, Nat. Commun. **13**, 5123 (2022).

[24] L. Peters et al., Millimetre-wave comb generated by an optical microcomb, Nat. Commun. **17**, 9833 (2026).

[25] J. Zang, T. C. Briles, J. S. Morgan, A. Beling, and S. B. Papp, Wide-band millimetre-wave synthesizer using microresonator-soliton photomixing, Nat. Photon. **20,** 750-757 (2026).

[26] C. W. Berry, N. Wang, M. R. Hashemi, M. Unlu, and M. Jarrahi, Significant performance enhancement in photoconductive terahertz optoelectronics by incorporating plasmonic contact electrodes, Nat. Commun. **4**, 1622 (2013).

[27] C. W. Berry, M. R. Hashemi, S. Preu, H. Lu, A. C. Gossard, and M. Jarrahi, High power terahertz generation using 1550 nm plasmonic photomixers, Appl. Phys. Lett. **105**, 011121 (2014).

[28] S.-H. Yang and M. Jarrahi, Frequency-tunable continuous-wave terahertz sources based on GaAs plasmonic photomixers, Appl. Phys. Lett. **107**, 131111 (2015).

[29] D. Turan, P. K. Lu, N. T. Yardimci, Z. Liu, L. Luo, J.-M. Park, U. Nandi, J. Wang, S. Preu, and M. Jarrahi, Wavelength conversion through plasmon-coupled surface states, Nat. Commun. **12**, 4641 (2021).

[30] M. Li, Y. Xia, Y. Chen, S. Tao, M. He, K. Huang, H. Li, M. Yan, and H. Zeng, Broadband terahertz comb with sub-Hz comb linewidth, Adv. Photonics **8**, 026015 (2026).

[31] T. J. Kippenberg, S. M. Spillane, and K. J. Vahala, Kerr-Nonlinearity Optical Parametric Oscillation in an Ultrahigh-Q Toroid Microcavity, Phys. Rev. Lett. **93**, 083904 (2004).

[32] A. B. Matsko, A. A. Savchenkov, D. Strekalov, V. S. Ilchenko, and L. Maleki, Optical hyperparametric oscillations in a whispering-gallery-mode resonator: Threshold and phase diffusion, Phys. Rev. A **71**, 033804 (2005).

[33] N. L. B. Sayson, T. Bi, V. Ng, H. Pham, L. S. Trainor, H. G. L. Schwefel, S. Coen, M. Erkintalo, and S. G. Murdoch, Octave-spanning tunable parametric oscillation in crystalline Kerr microresonators, Nat. Photonics **13**, 701 (2019).

[34] J. R. Stone, G. Moille, X. Lu, and K. Srinivasan, Conversion Efficiency in Kerr-Microresonator Optical Parametric Oscillators: From Three Modes to Many Modes, Phys. Rev. Appl. **17**, 024038 (2022).

[35] E. F. Perez, G. Moille, X. Lu, J. Stone, F. Zhou, and K. Srinivasan, High-performance Kerr microresonator optical parametric oscillator on a silicon chip, Nat. Commun. **14**, 242 (2023).

[36] X. Xue, Y. Xuan, P. Wang, Y. Liu, D. E. Leaird, M. Qi, and A. M. Weiner, Normal-dispersion microcombs enabled by controllable mode interactions, Laser Photonics Rev. **9**, L23 (2015).

[37] C. M. Gentry, X. Zeng, and M. A. Popović, Tunable coupled-mode dispersion compensation and its application to on-chip resonant four-wave mixing, Opt. Lett. **39**, 5689 (2014).

[38] X. Xue, X. Zheng, and B. Zhou, Super-efficient temporal solitons in mutually coupled optical cavities, Nat. Photonics **13**, 616 (2019).

[39] Ó. B. Helgason, F. R. Arteaga-Sierra, Z. Ye, K. Twayana, P. A. Andrekson, M. Karlsson, J. Schröder, and V. Torres-Company, Dissipative solitons in photonic molecules, Nat. Photonics **15**, 305 (2021).

[40] A. Tikan et al., Emergent nonlinear phenomena in a driven dissipative photonic dimer, Nat. Phys. **17**, 604 (2021).

[41] D. Pidgayko, A. Tusnin, J. Riemensberger, A. Stroganov, A. Tikan, and T. J. Kippenberg, Voltage-tunable optical parametric oscillator with an alternating dispersion dimer integrated on a chip, Optica **10**, 1582 (2023).

[42] Ó. B. Helgason, M. Girardi, Z. Ye, F. Lei, J. Schröder, and V. Torres-Company, Surpassing the nonlinear conversion efficiency of soliton microcombs, Nat. Photonics **17**, 992 (2023).

[43] J. R. Stone, T. C. Briles, T. E. Drake, D. T. Spencer, D. R. Carlson, S. A. Diddams, and S. B. Papp, Thermal and Nonlinear Dissipative-Soliton Dynamics in Kerr-Microresonator Frequency Combs, Phys. Rev. Lett. **121**, 063902 (2018).

[44] T. E. Drake, J. R. Stone, T. C. Briles, and S. B. Papp, Thermal decoherence and laser cooling of Kerr microresonator solitons, Nat. Photonics **14**, 480 (2020).

[45] G. Marty, S. Combrié, F. Raineri, and A. De Rossi, Photonic crystal optical parametric oscillator, Nat. Photonics **15**, 53 (2021).

[46] J. A. Black, G. Brodnik, H. Liu, S.-P. Yu, D. R. Carlson, J. Zang, T. C. Briles, and S. B. Papp, Optical-parametric oscillation in photonic-crystal ring resonators, Optica **9**, 1183 (2022).

[47] J. R. Stone, X. Lu, G. Moille, and K. Srinivasan, Efficient chip-based optical parametric oscillators from 590 to 1150 nm, APL Photonics **7**, 121301 (2022).

[48] F. Lei, Z. Ye, K. Twayana, Y. Gao, M. Girardi, Ó. B. Helgason, P. Zhao, and V. Torres-Company, Hyperparametric Oscillation via Bound States in the Continuum, Phys. Rev. Lett. **130**, 093801 (2023).

[49] H. Liu, G. M. Brodnik, J. Zang, D. R. Carlson, J. A. Black, and S. B. Papp, Threshold and Laser Conversion in Nanostructured-Resonator Parametric Oscillators, Phys. Rev. Lett. **132**, 023801 (2024).

[50] J. Li, Y. Zhang, J. Zeng, and S. Yu, Broadband and accurate electric tuning of on-chip efficient nonlinear parametric conversion, Optica **12**, 424 (2025).

[51] Y. Sun, F. Lei, Y. Gao, and V. Torres-Company, High-power on-chip hyperparametric oscillator, Opt. Lett. **50**, 4798 (2025).

[52] E. A. Kittlaus, D. Eliyahu, S. Ganji, S. Williams, A. B. Matsko, K. B. Cooper, and S. Forouhar, A low-noise photonic heterodyne synthesizer and its application to millimeter-wave radar, Nat. Commun. **12**, 4397 (2021).

[53] L. Djevahirdjian, L. Lechevallier, M.-A. Martin-Drumel, O. Pirali, G. Ducournau, R. Kassi, and S. Kassi, Frequency stable and low phase noise THz synthesis for precision spectroscopy, Nat. Commun. **14**, 7162 (2023).

[54] J. R. Stone and S. B. Papp, Harnessing Dispersion in Soliton Microcombs to Mitigate Thermal Noise, Phys. Rev. Lett. **125**, 153901 (2020).

[55] G. D. Domenico, S. Schilt, and P. Thomann, Simple approach to the relation between laser frequency noise and laser line shape, Appl. Opt. **49**, 4801 (2010).

[56] Y. Zhao, S.-E. Zumrat, S.-A. Tsao, and M. Jarrahi, Terahertz generation and detection through gain-enhanced interband photomixing in quantum well structures, Nat. Commun. **17**, 6247 (2026).

[57] J. A. Barnes et al., Characterization of Frequency Stability, IEEE Trans. Instrum. Meas. **IM–20**, 105 (1971).

**Author Contributions.**

D.I.L. and F.S. designed and led the experimental study, conducted the measurements, analyzed the data, and prepared the manuscript. W.L. contributed to data analysis and interpretation. X.J. assisted with the construction of the initial experimental setup. A.T. contributed to device fabrication. A.A. contributed to the initial device concept. M.C. and J.L. contributed to the experiments and data analysis. M.J. and C.W.W. supervised the project and revised the manuscript with input from all authors. All authors reviewed and approved the final manuscript. D.I.L. and F.S. contributed equally to this work.

**Data availability**

The data that support the plots within this paper and other findings of this study are available from the corresponding author upon request.

**Competing interests**

M.J. is a co-founder of Lookin Inc. The other authors do not have conflicts of interest related to the contents of this manuscript.

**Acknowledgment**

The authors acknowledge support from the Defense Advanced Research Projects Agency (DARPA) under the EQSTRA program (grant # HR0011-25-2-0006) and the Airforce Office of Scientific Research (grant #FA9550-25-1-0358). Development of the plasmonic photomixers used in this work was supported by the Department of Energy (grant # DE-SC0016925).

**Figure 1 | Concept of a photonic multi-tone mmWave/THz source. a,** Conceptual architecture of the dual-ring Kerr OPO followed by broadband photomixing of the generated multi-line optical spectrum. **a(i),** Dual-ring resonator and AMX-enabled OPO concept, illustrating the measured integrated dispersion $D_{int}/2\pi$, intracavity-field redistribution, and hybrid-mode splitting $\delta_{rel}$ near 1566.38 nm. $\lambda_p$ and $\lambda_{AMX}$ represent the potential pump wavelength and the hybridization resonance wavelength of the dual-ring system, respectively. **a(ii),** Broadband photomixing, in which pairwise optical frequency differences, $\Delta\nu_{mn}= |\nu_m-\nu_n|$, are simultaneously converted into mmWave/THz carriers. A photomixer incorporating plasmonic contact electrodes is used to enhance the optical-to-mmWave/THz conversion efficiency. The electrode geometry is designed to excite surface plasmon waves over the OPO wavelength range, confining photogenerated carriers in close proximity to the plasmonic electrodes. Numerical simulations of optical absorption (Lumerical) indicate that approximately 29% of the optical power in the OPO lines is absorbed within 100 nm of the plasmonic contact electrodes, enabling efficient mmWave/THz generation. **b**, Conceptual multi-band transmitter architecture incorporating optical demultiplexing, modulation, multiplexing, phase control, and parallel photomixing to generate simultaneous mmWave/THz communication channels.

**Figure 2 | Reconfigurable multi-tone mmWave/THz generation. a,** Experimental setup for direct photomixing and harmonic down-conversion of the dual-ring OPO output. **b,** Optical spectra exhibiting OPO frequency spacings of 400, 300, 200, and 100 GHz. **c,** Corresponding down-converted mmWave/THz spectra. Different frequency bands were measured sequentially using the appropriate harmonic mixers, while all tones associated with each OPO state were generated

simultaneously. **d,** Representative optical microscopy images of the dual-ring resonator and broadband photomixer.

**Figure 3 | High-efficiency OPO and hot-cavity characterization. a,** Experimental configuration for detuning-dependent OPO and counter-propagating hot-cavity measurements. **b,** Evolution of the 400-GHz OPO spectrum with pump detuning; dashed lines indicate the operating points corresponding to maximum conversion efficiency (i) and maximum pump depletion (ii). **c,** Hot-cavity transmission near the targeted OPO wavelength, with an extracted mode splitting of 168.4–190.9 MHz; insets show the corresponding optical field distributions in the dual-ring resonator and an optical microscopy image of the device. **d,** Total and primary signal-idler pair conversion efficiencies as functions of pump detuning. **e,** Representative OPO states with frequency separations ranging from 100 GHz to 1.2 THz. **f,** Comparison of primary-sidebands conversion efficiency ($\eta_{pri}$) versus pump-to-primary-sideband frequency separation achieved in this work, with those of previously reported integrated Kerr optical parametric oscillators [45–51].

**Figure 4 | Broadband and continuously tunable mmWave/THz generation. a,** Experimental configurations for (i) harmonic down-conversion and individual-carrier characterization and (ii) radiated-power measurements under modulated optical excitation. **b,** Down-converted spectra of the generated carriers spanning 100 GHz to 1.1 THz. **c,** Continuous carrier-frequency tuning achieved by varying the pump detuning for different OPO states. **d,** Measured radiated power as a function of carrier frequency under modulated excitation with a 1% duty cycle, exceeding 2 mW at 100 GHz.

**Figure 5| Frequency-noise characteristics and long-term stability. a,** Frequency-noise PSDs of the 300-, 600-, and 700-GHz carriers, together with the β-separation line used for linewidth estimation. **b,** Allan deviation of the 800-GHz carrier, calculated from phase-noise measurements at short averaging times and obtained from direct frequency-counter measurements at longer averaging times. **c,** Long-term frequency fluctuations of the free-running 800-GHz carrier measured over approximately 30 min.

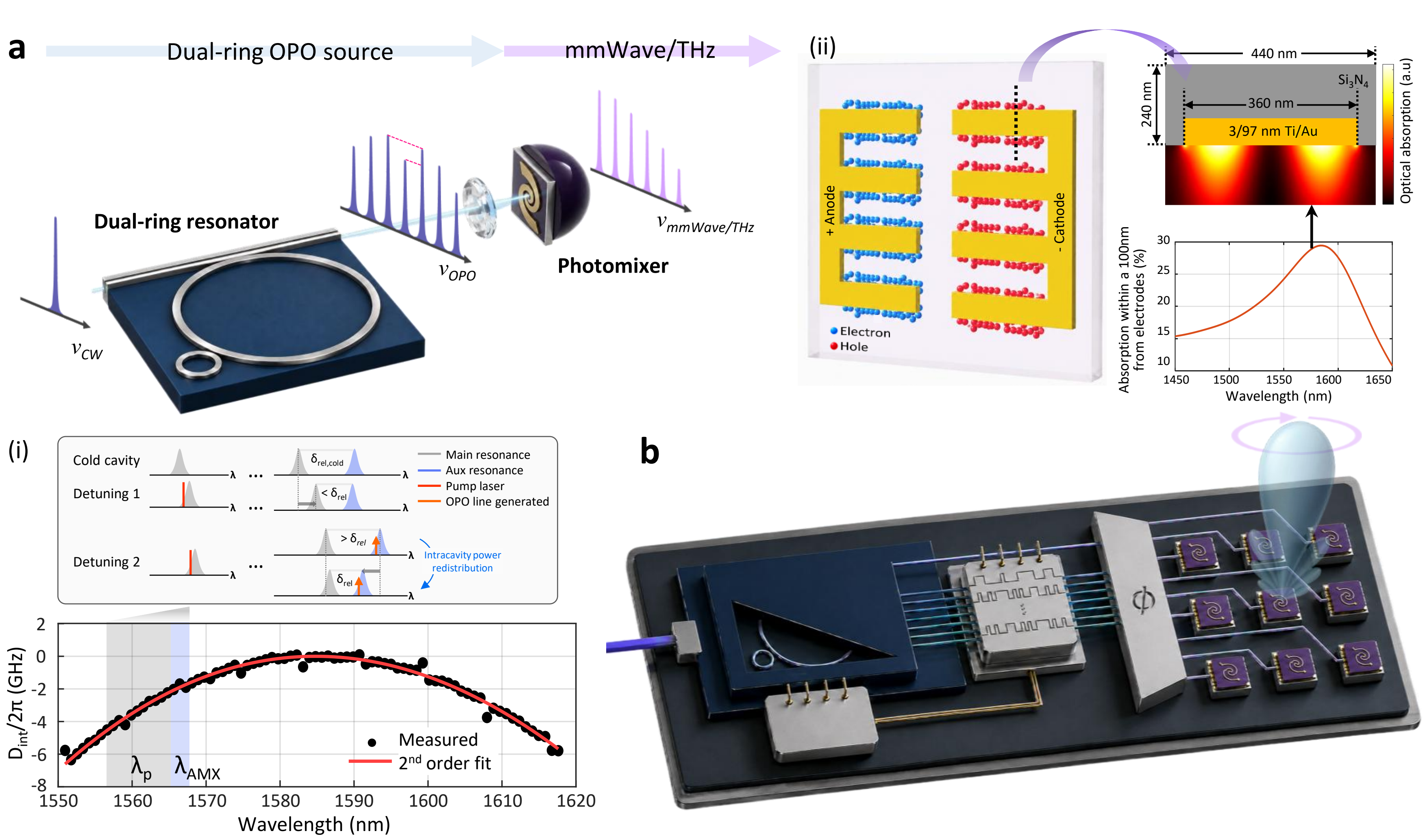

a
Dual-ring OPO source
mmWave/THz
(ii)
Dual-ring resonator
$\nu_{CW}$
$\nu_{OPO}$
Photomixer
$\nu_{mmWave/THz}$
+ Anode
- Cathode
Electron
Hole
440 nm
$Si_3N_4$
360 nm
240 nm
3/97 nm Ti/Au
Optical absorption (a.u)
Absorption within a 100nm from electrodes (%)
Wavelength (nm)
(i)
Cold cavity
Detuning 1
Detuning 2
$\delta_{rel,cold}$
$< \delta_{rel}$
$> \delta_{rel}$
$\delta_{rel}$
Main resonance
Aux resonance
Pump laser
OPO line generated
Intracavity power redistribution
$D_{int}/2\pi$ (GHz)
$\lambda_p$
$\lambda_{AMX}$
Measured
2nd order fit
Wavelength (nm)
b

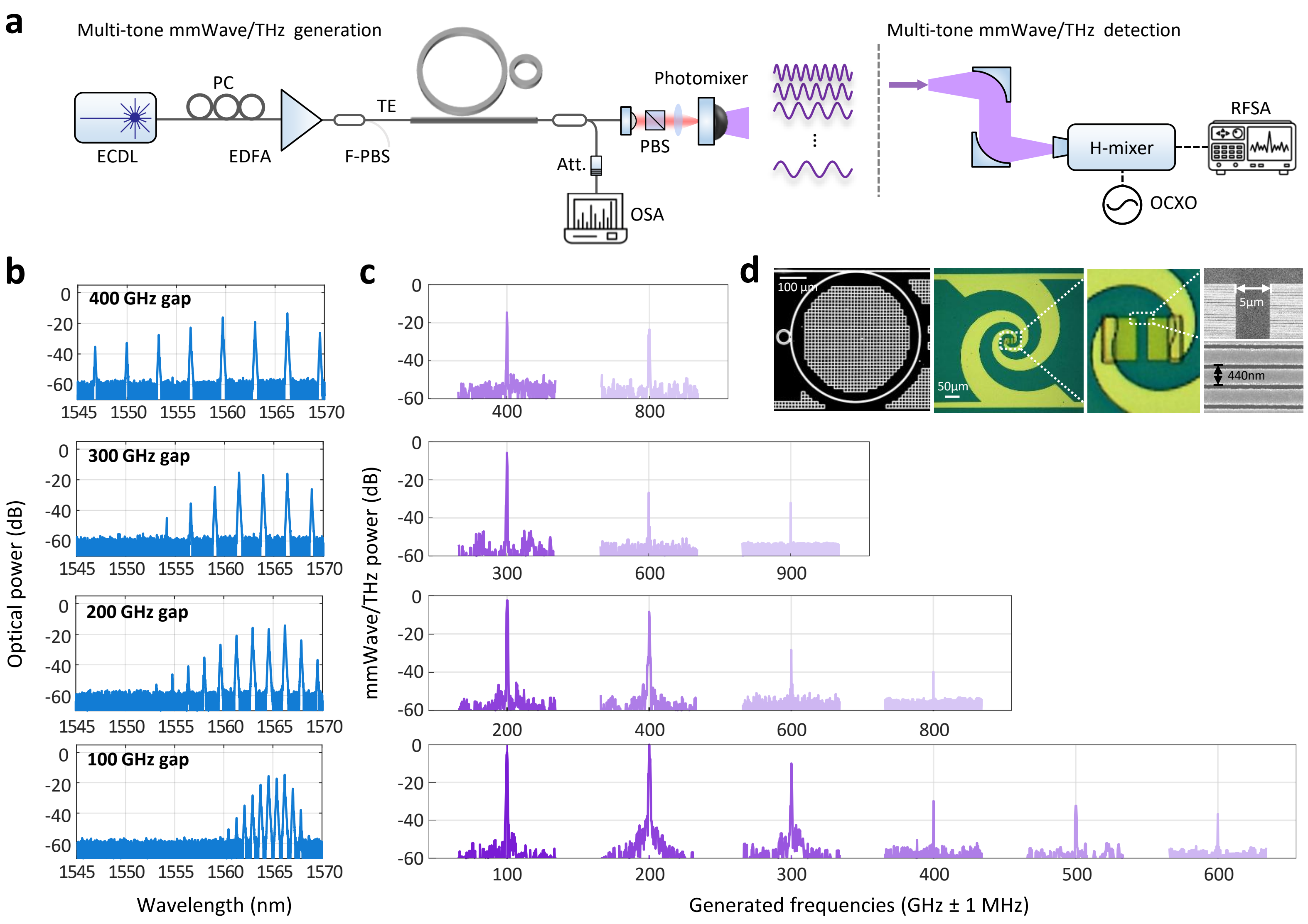
a
Multi-tone mmWave/THz generation
Multi-tone mmWave/THz detection
PC
ECDL
EDFA
TE
F-PBS
Att.
OSA
Photomixer
PBS
H-mixer
RFSA
OCXO
b
400 GHz gap
300 GHz gap
200 GHz gap
100 GHz gap
0
-20
-40
-60
1545 1550 1555 1560 1565 1570
Optical power (dB)
Wavelength (nm)
c
400
800
300
600
900
200
400
600
800
100
200
300
400
500
600
mmWave/THz power (dB)
Generated frequencies (GHz ± 1 MHz)
d
100 µm
50µm
5µm
440nm

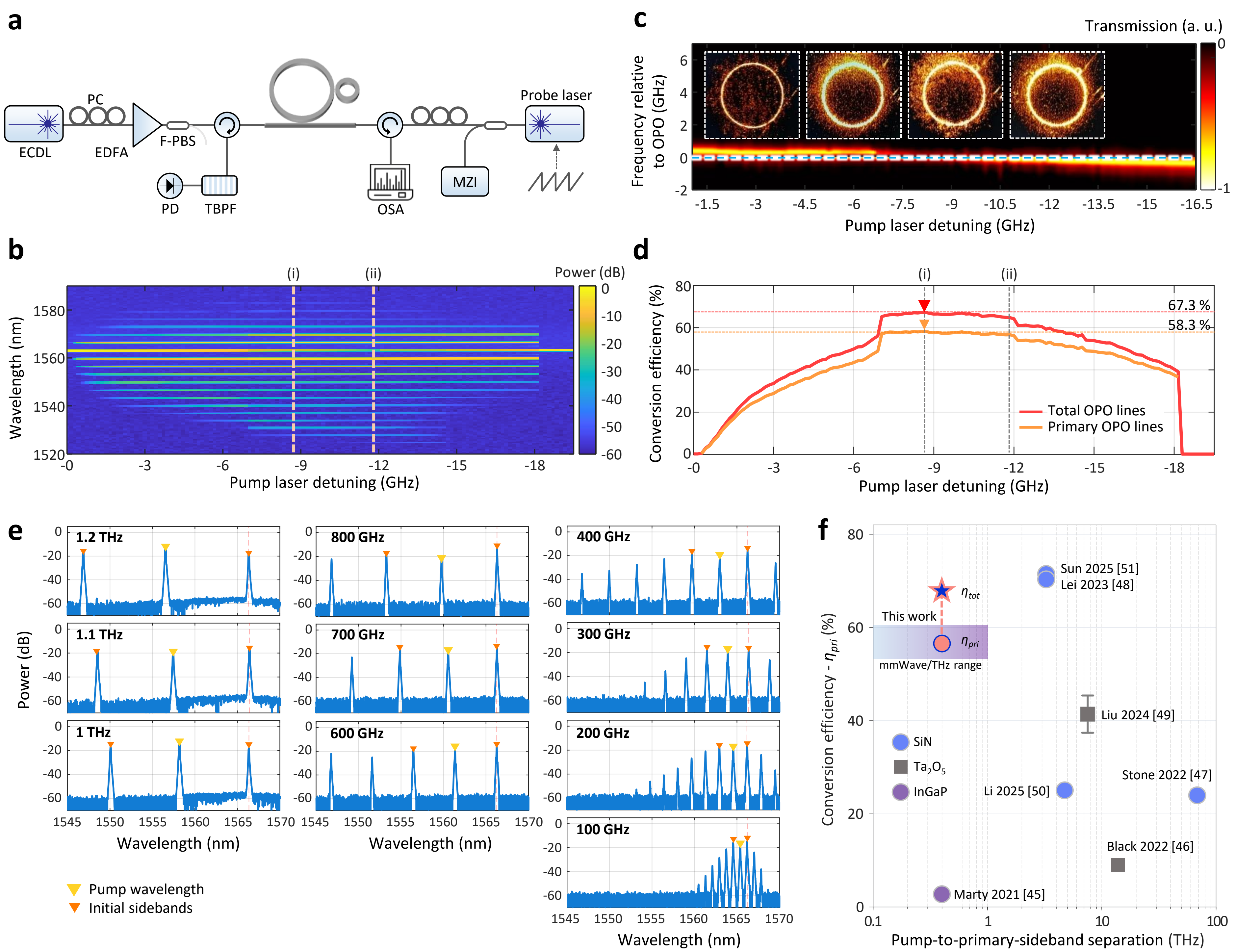

a
PC
ECDL
EDFA
F-PBS
PD
TBPF
OSA
MZI
Probe laser
b
(i)
(ii)
Power (dB)
Wavelength (nm)
Pump laser detuning (GHz)
c
Transmission (a. u.)
Frequency relative to OPO (GHz)
Pump laser detuning (GHz)
d
Conversion efficiency (%)
67.3 %
58.3 %
Total OPO lines
Primary OPO lines
Pump laser detuning (GHz)
e
1.2 THz
800 GHz
400 GHz
1.1 THz
700 GHz
300 GHz
1 THz
600 GHz
200 GHz
100 GHz
Power (dB)
Wavelength (nm)
Pump wavelength
Initial sidebands
f
Conversion efficiency - $\eta_{pri}$ (%)
Sun 2025 [51]
Lei 2023 [48]
$\eta_{tot}$
This work
$\eta_{pri}$
mmWave/THz range
Liu 2024 [49]
SiN
$Ta_2O_5$
InGaP
Li 2025 [50]
Stone 2022 [47]
Black 2022 [46]
Marty 2021 [45]
Pump-to-primary-sideband separation (THz)

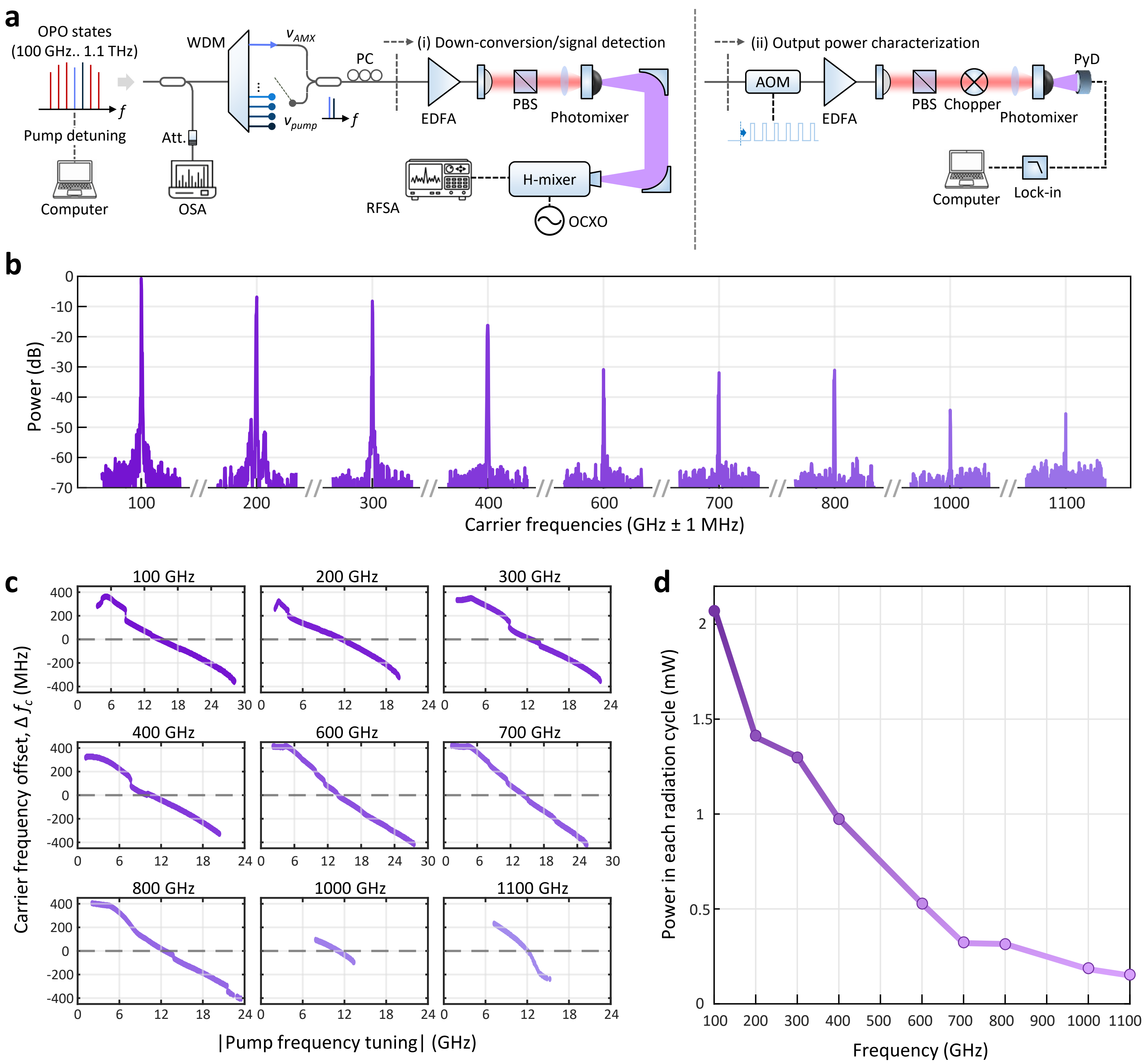
a
OPO states
(100 GHz.. 1.1 THz)
Pump detuning
Computer
WDM
Att.
OSA
PC
(i) Down-conversion/signal detection
EDFA
PBS
Photomixer
RFSA
H-mixer
OCXO
(ii) Output power characterization
AOM
EDFA
PBS
Chopper
Photomixer
PyD
Computer
Lock-in
b
Power (dB)
Carrier frequencies (GHz ± 1 MHz)
c
100 GHz
200 GHz
300 GHz
400 GHz
600 GHz
700 GHz
800 GHz
1000 GHz
1100 GHz
Carrier frequency offset, Δ $f_c$ (MHz)
|Pump frequency tuning| (GHz)
d
Power in each radiation cycle (mW)
Frequency (GHz)

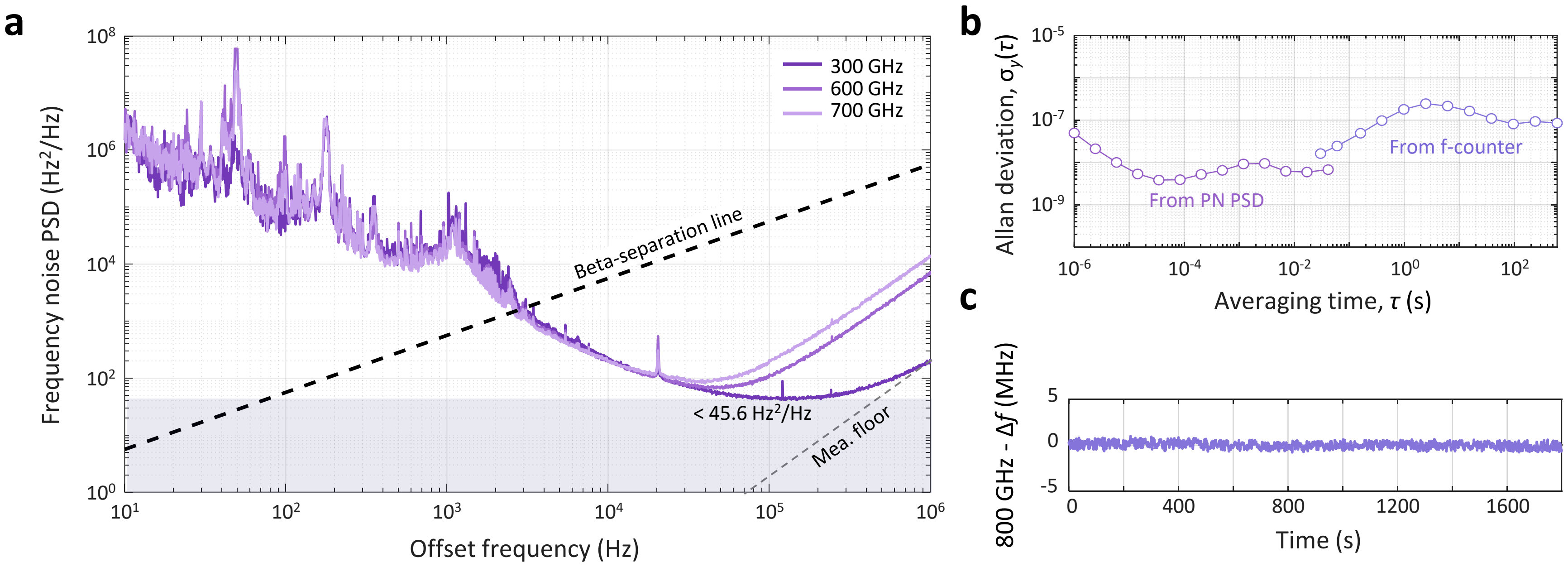

a
300 GHz
600 GHz
700 GHz
Beta-separation line
< 45.6 Hz²/Hz
Mea. floor
Frequency noise PSD (Hz²/Hz)
Offset frequency (Hz)
b
Allan deviation, σy(τ)
From f-counter
From PN PSD
Averaging time, τ (s)
c
800 GHz - Δf (MHz)
Time (s)

**Supplementary Information**

# Reconfigurable multi-tone mmWave/THz generation powered by high-efficiency Kerr parametric oscillation

Dong IL Lee[1,†,*], Fanqi Shen[2,†], Wenzheng Liu[1], Xinghe Jiang[2], Anika Tabassum[2], Alwaleed Aldhafeeri[1], Manas Choure[1], Jasmine Lu[1], Mona Jarrahi[2,*], and Chee Wei Wong[1,*]

[1]Fang Lu Mesoscopic Optics and Quantum Electronics Laboratory, University of California, Los Angeles, CA 90095, USA

[2]Terahertz Electronics Laboratory, University of California, Los Angeles, CA 90095, USA.

[†] These authors contributed equally to this work

[*]Correspondence to: dilee@ucla.edu mjarrahi@ucla.edu, cheewei.wong@ucla.edu

## Supplementary Information Content

## I. Dual-ring resonator design and cold-cavity characterization

The dual-ring resonator is fabricated on a 350-nm-thick silicon nitride (SiN) platform. The main and auxiliary rings have radii of 238.15 and 23.79 µm and waveguide widths of 2.0 and 3.5 µm, respectively, with an inter-ring coupling gap of 350 nm. The corresponding free spectral ranges (FSRs) are 100.53 GHz and approximately 1 THz, respectively. Figure S1a presents the simulated modal dispersion as a function of wavelength and waveguide width. The selected waveguide widths, indicated by the red dashed lines, yield normal dispersion with $D_2/2\pi$ values of -6.31 MHz and -767.89 MHz for the main and auxiliary rings, respectively, over the target wavelength range. The substantially larger FSR of the auxiliary ring results in a sparse resonance spectrum, enabling localized perturbations of the more densely spaced resonances of the main ring through inter-ring coupling.

The cold-cavity resonance frequencies are characterized using the swept-wavelength interferometry (SWI) setup shown in Fig. S1b. The output of a tunable external-cavity diode laser (ECDL) is split into three paths for device transmission, interferometric calibration, and gas-cell referencing. Polarization controllers and a fiber polarization beam splitter are used to control the input polarization and selectively excite the fundamental transverse-electric ($TE_{00}$) mode. A Mach–Zehnder interferometer (MZI) with a 50-m fiber delay provides relative frequency calibration, while the absorption features of hydrogen cyanide serve as absolute frequency references. The device transmission, interferometer signal, and gas-cell transmission are simultaneously recorded using an FPGA-based data acquisition system, enabling accurate extraction of the cavity resonance frequencies on a calibrated frequency axis. For the cold-cavity measurements presented in Figs. S1c and S1d, the optical power measured immediately before the chip is −18 dBm.

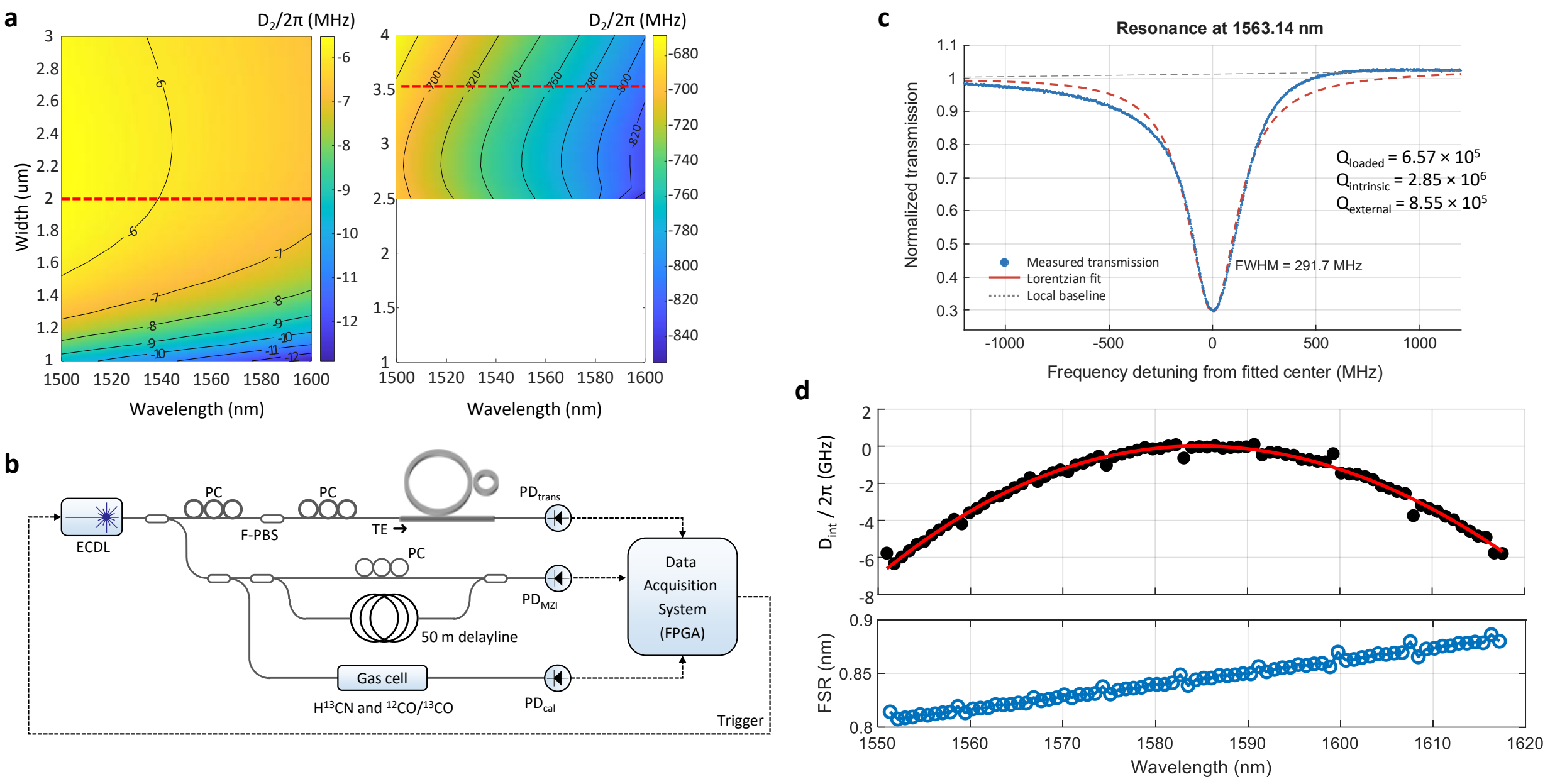

**Figure S1 | Design and cold-cavity characterization of the dual-ring resonator. a**, Simulated dispersion maps of the main and auxiliary rings. Red dashed lines indicate the selected waveguide widths of 2.0 and 3.5 μm, respectively. **b**, Swept-wavelength interferometry setup for calibrated cavity-transmission measurements. **c**, Normalized transmission spectrum of a resonance near 1563.14 nm, together with a Lorentzian fit yielding a linewidth of 291.7 MHz and a loaded quality factor of $Q_{loaded} = 6.57\times10^5$. **d**, Measured integrated dispersion and corresponding second-order fit (top), and free spectral range (FSR) expressed in wavelength units (bottom). The fitted second-order dispersion coefficient of $D_2/2\pi = -7.86$ MHz confirms normal dispersion.

Figure S1c shows the normalized transmission spectrum of a representative resonance near 1563.14 nm. A Lorentzian fit accounting for the local baseline yields a full width at half maximum (FWHM) of 291.7 MHz and a loaded quality factor of $Q_{Loaded} = 6.57 \times 10^5$. The fit further yields intrinsic and external coupling quality factors of $Q_{intrinsic} = 2.85 \times 10^6$ and $Q_{external} = 8.55 \times 10^5$, respectively. The lower external coupling quality factor indicates that coupling to the bus waveguide contributes more strongly to the loaded linewidth than intrinsic cavity losses, consistent with operation in the overcoupled regime.

The calibrated resonance frequencies are used to extract the integrated dispersion shown in Fig. S1d. A second-order fit reveals a smooth normal-dispersion background, while localized deviations from the fit indicate mode interactions, including the targeted avoided mode crossing near 1566.38 nm. The lower panel shows the corresponding spacing between adjacent resonances, expressed in wavelength units, where periodic hybridization between the main- and auxiliary-ring modes is evident. These cold-cavity measurements establish a reference for the coupled-mode and hot-cavity analyses presented in Supplementary Note 2.

## II. Hot-cavity resonance characterization near an avoided mode crossing

To characterize the local resonance response under optical parametric oscillation OPO conditions, the hot-cavity transmission near the OPO wavelength of 1566.38 nm is measured using a counter-propagating probe laser. The probe-frequency axis is calibrated using an MZI with a fringe spacing of 3.95 MHz. The local main–auxiliary resonance splitting, $\delta_{rel}$, is defined as the frequency separation between the two transmission minima of the coupled-ring response throughout this work. A representative probe-transmission spectrum and the simulated inter-ring coupling response are presented in Fig. S2.1.

The measured transmission spectrum in Fig. S2.1a is fitted using a steady-state two-resonator coupled-mode model [S1]. For a bus waveguide directly coupled to the main ring, the normalized power transmission is given by:

$$T(\omega) = \left| 1 - \frac{\kappa_{ex}}{\frac{\kappa_m}{2} - i(\omega - \omega_m) + \frac{g^2}{\frac{\kappa_a}{2} - i(\omega - \omega_a)}} \right|^2$$

Here, $T$ denotes the normalized through-port power transmission, defined as the ratio of the optical power at the through-port output to the input optical power. The probe angular frequency is $\omega$, while $\omega_m$ and $\omega_a$ represent the effective uncoupled resonance angular frequencies of the main and auxiliary rings, respectively, under the operating conditions. The parameter $g$ denotes the coherent inter-ring coupling rate. The quantities $\kappa_m$ and $\kappa_a$ represent the total energy decay rates of the main- and auxiliary-ring modes, respectively, while $\kappa_{ex}$ denotes the main-ring energy decay rate associated with coupling to the bus waveguide. All angular frequencies and the coupling rate, $g$, are expressed in rad/s, whereas the energy decay rates are expressed in $s^{-1}$. An overall amplitude scaling factor and a linear background term are included when fitting the model to the measured signal.

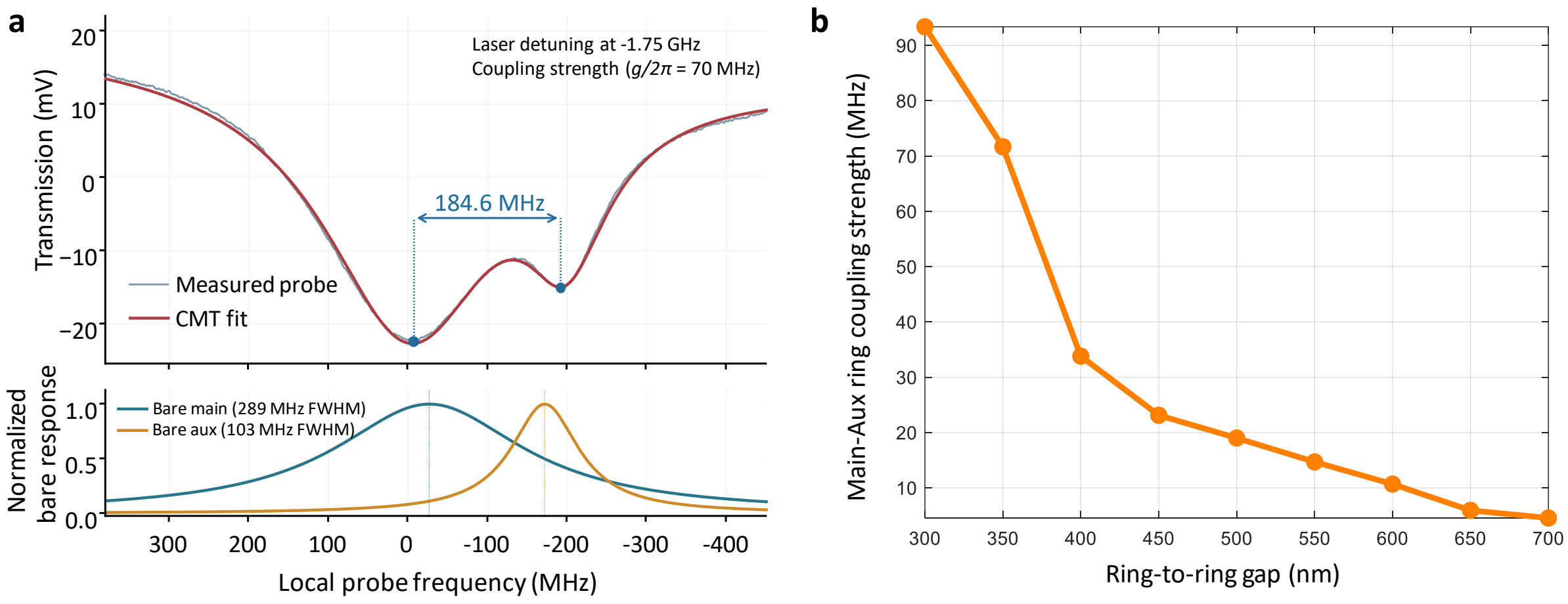

**Figure S2.1 | Hot-cavity resonance characterization and inter-ring coupling. a,** Measured probe-transmission signal (gray) and coupled-mode theory (CMT) fit (red) at a pump-frequency detuning of −1.75 GHz, with $g/(2\pi) = 70$ MHz. Blue markers indicate the two measured transmission minima, separated by $\delta_{rel} = 184.6$ MHz. The lower panel shows the normalized uncoupled-ring responses inferred from the fit, with linewidths of 289 MHz and 103 MHz for the main and auxiliary rings, respectively. **b,** Inter-ring coupling obtained from finite-difference time-domain (FDTD) simulations as a function of the ring-to-ring gap. The simulated coupling is approximately 70 MHz at a gap of 350 nm.

FDTD simulations for a 350-nm ring-to-ring gap estimate an inter-ring coupling rate of approximately $g/2\pi = 70$ MHz (Fig. S2.1b). This value also yields the lowest fitting cost among the coupling values tested against spectra with two resolved minima. Accordingly, $g/2\pi$ is held fixed at 70 MHz throughout the analysis, with weak regularization applied to the main-ring linewidth around $\kappa_m/2\pi = 290$ MHz. The representative fit reproduces the positions of both transmission minima (Fig. S2.1a). Across the resolved spectra used for validation, independent full-trace fits reproduce the directly measured separation with a mean absolute difference of approximately 2 MHz.

During OPO operation, strong emission is accompanied by a narrow, spike-like feature in the probe transmission. We attribute this feature to heterodyne beating between the counter-propagating probe and backscattered OPO light at the detector, consistent with probe–comb beating observed in hot-cavity spectroscopy [S2]. This contribution becomes particularly distinct above a pump-laser detuning of approximately −7 GHz in the main-text hot-cavity scan (Fig. 3c), distorting the measured transmission profile of one of the hybridized resonances. The resulting overlap makes direct identification of the transmission minimum less reliable and leads us to reconstruct the underlying coupled-ring response.

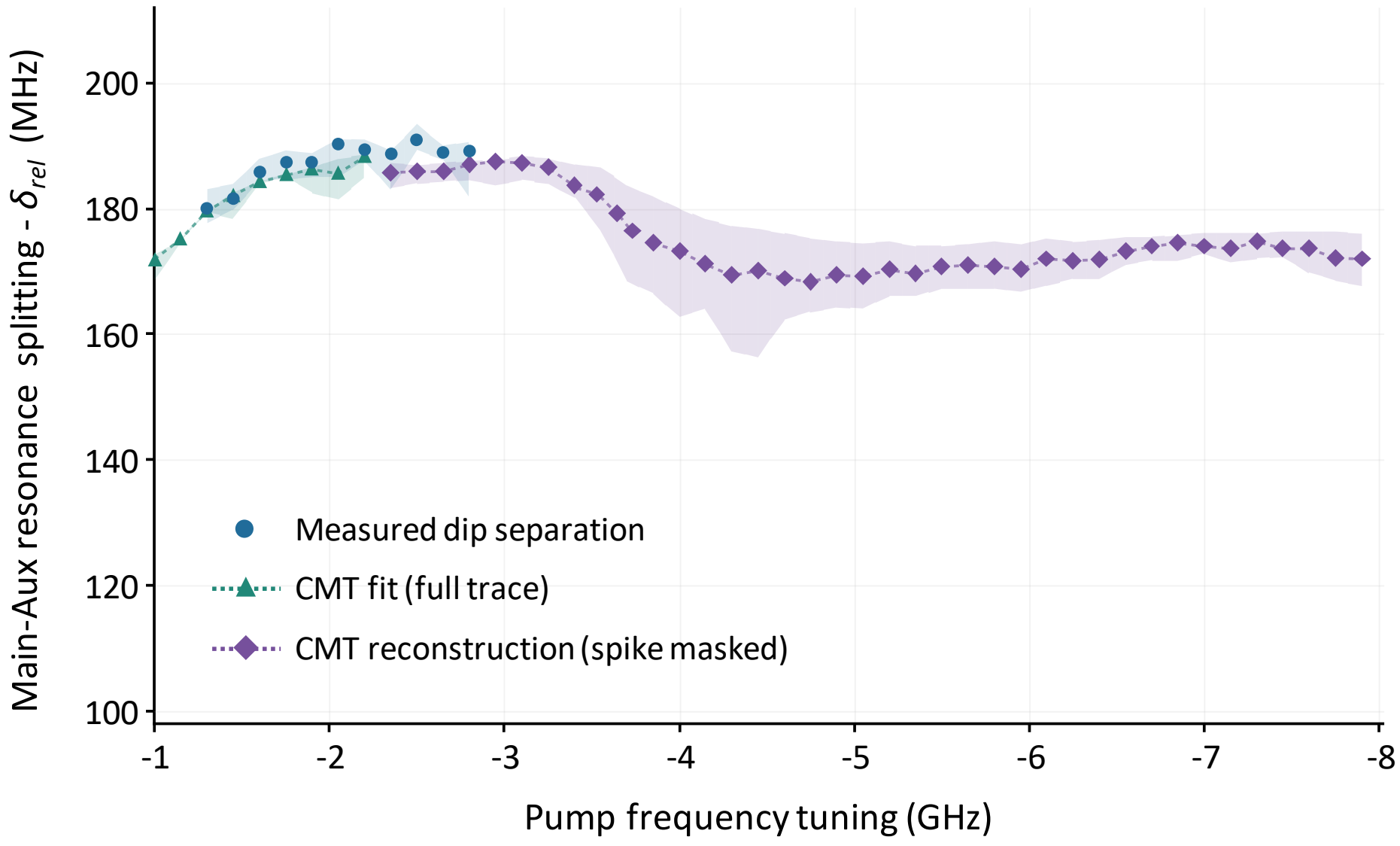

**Figure S2.2 | Hybridized resonance separation during pump-frequency tuning.** Directly measured dip separations are compared with those obtained from full-trace and spike-masked CMT analyses. The extracted median separation initially increases and then decreases toward a plateau, remaining within 168.4–190.9 MHz over the plotted tuning range. Shaded bands indicate the corresponding minimum-to-maximum ranges.

To extract the dip separation across the tuning range, full-trace fitting is used for the initial spectra, while masked fitting is applied to traces affected by the heterodyne beat feature. Each initial trace is fitted independently, without an exclusion window or a continuity constraint between adjacent pump settings. For the affected traces, a ±60-MHz window around the feature is excluded from the fit, and parameter variations between adjacent pump settings are constrained. The full transmission trace is then reconstructed across the excluded region, retaining the fitted scale factor and linear background. The splitting is extracted from the reconstructed trace only when it exhibits two distinct minima. Validation using resolved spectra with one minimum artificially masked reveals an average underestimation of approximately 4–5 MHz in the continuity-assisted reconstruction.

Figure S2.2 summarizes the evolution of $\delta_{rel}$. At each pump setting, markers for direct measurements and unmasked CMT fits denote the median of four separation values obtained from repeated probe scans, with shaded bands indicating their minimum-to-maximum range. For masked CMT reconstructions, the same four scans are each fitted using two strengths of the continuity constraint between adjacent pump settings, while the coupling rate and mask width remain fixed. The corresponding markers and shaded bands represent the median and minimum-to-maximum range, respectively. These masked bands therefore account for both scan-to-scan uncertainty and sensitivity to the continuity constraint.

The extracted separation exhibits a nonmonotonic evolution over a range of 22.5 MHz (168.4–190.9 MHz), with an initial increase followed by intervals of comparatively weak variation separated by a decrease. Coupled-ring hybridization provides a mechanism for locally modifying cavity resonances and modal dispersion in normal-dispersion resonators [S3–S5]. Together with the continued oscillation of the same signal–idler pair, the observed evolution is consistent with a sustained local AMX-induced resonance.

## III. Representative OPO spectra exhibiting strong pump depletion

Figure S3a presents the optical spectrum of the 400-GHz-spaced OPO state corresponding to the 25.06-dB pump depletion reported in the main text. The spectrum shows a strongly suppressed residual pump, accompanied by strong primary signal and idler lines and additional sidebands on the same frequency grid.

Figure S3b provides an additional example at an OPO spacing of 1.1 THz, obtained by pumping near the next Vernier-induced mode-interaction region. This state exhibits a pump depletion of 17.18 dB, with the primary signal and idler lines likewise dominating over the residual pump. This observation demonstrates that substantial pump depletion is also achievable at a larger

OPO spacing by pumping near a neighboring mode-interaction region at 1560 nm. The simultaneous presence of strong pump suppression and prominent generated sidebands is consistent with efficient parametric power transfer in this additional operating region.

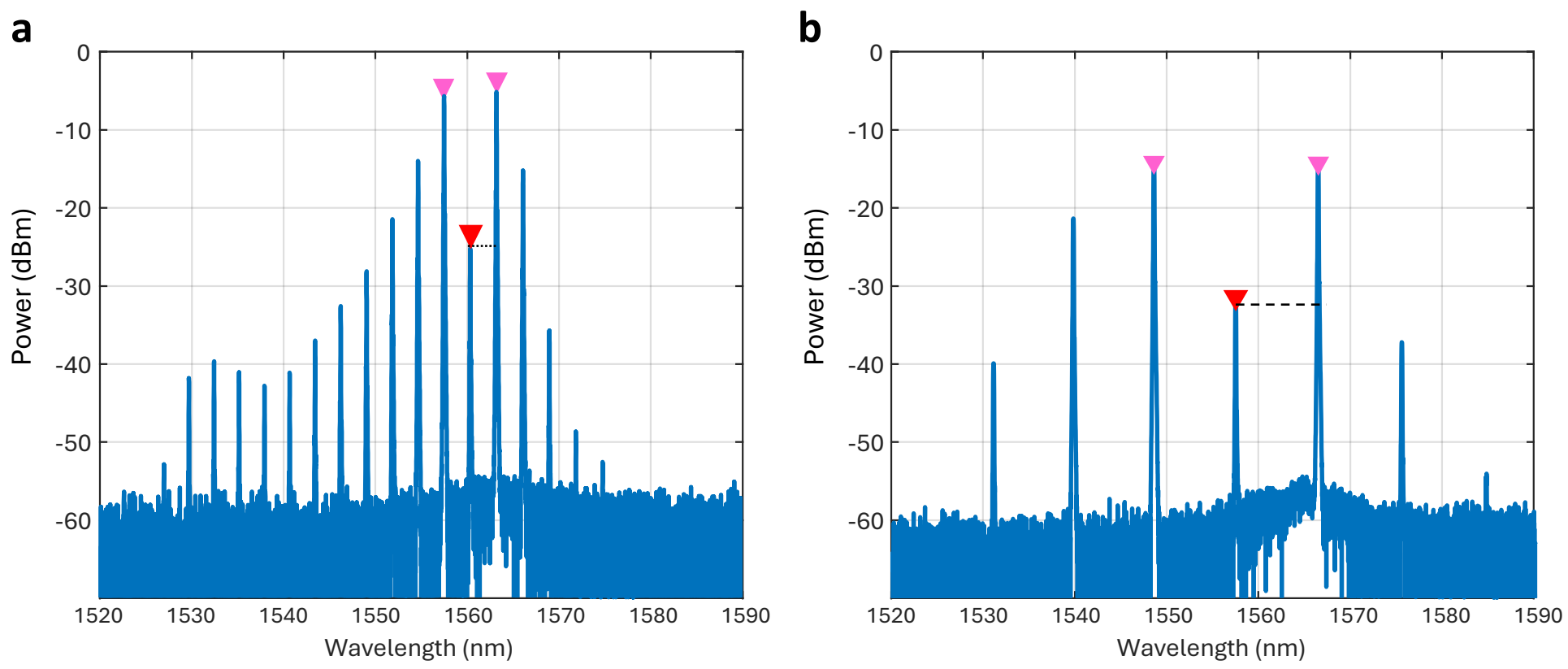

**Figure S3 | Representative OPO spectra with strong pump depletion. a**, 400-GHz-spaced OPO state exhibiting a pump depletion of 25.06 dB. **b**, 1.1-THz-spaced OPO state exhibiting a pump depletion of 17.18 dB. Red and magenta inverted triangles indicate the residual pump and primary signal–idler pair, respectively. The spectra in (a) and (b) were recorded with 15 and 25 dB of optical attenuation, respectively, before the optical spectrum analyzer (OSA) to protect the instrument.

## IV. Pulsed mmWave/THz radiation cycle

Figure S4.1 shows the dark current of the photomixer measured under continuous DC bias in the absence of optical illumination. The dark current increases substantially with increasing applied bias voltage, resulting in an increased power dissipation and thermal loading of the device.

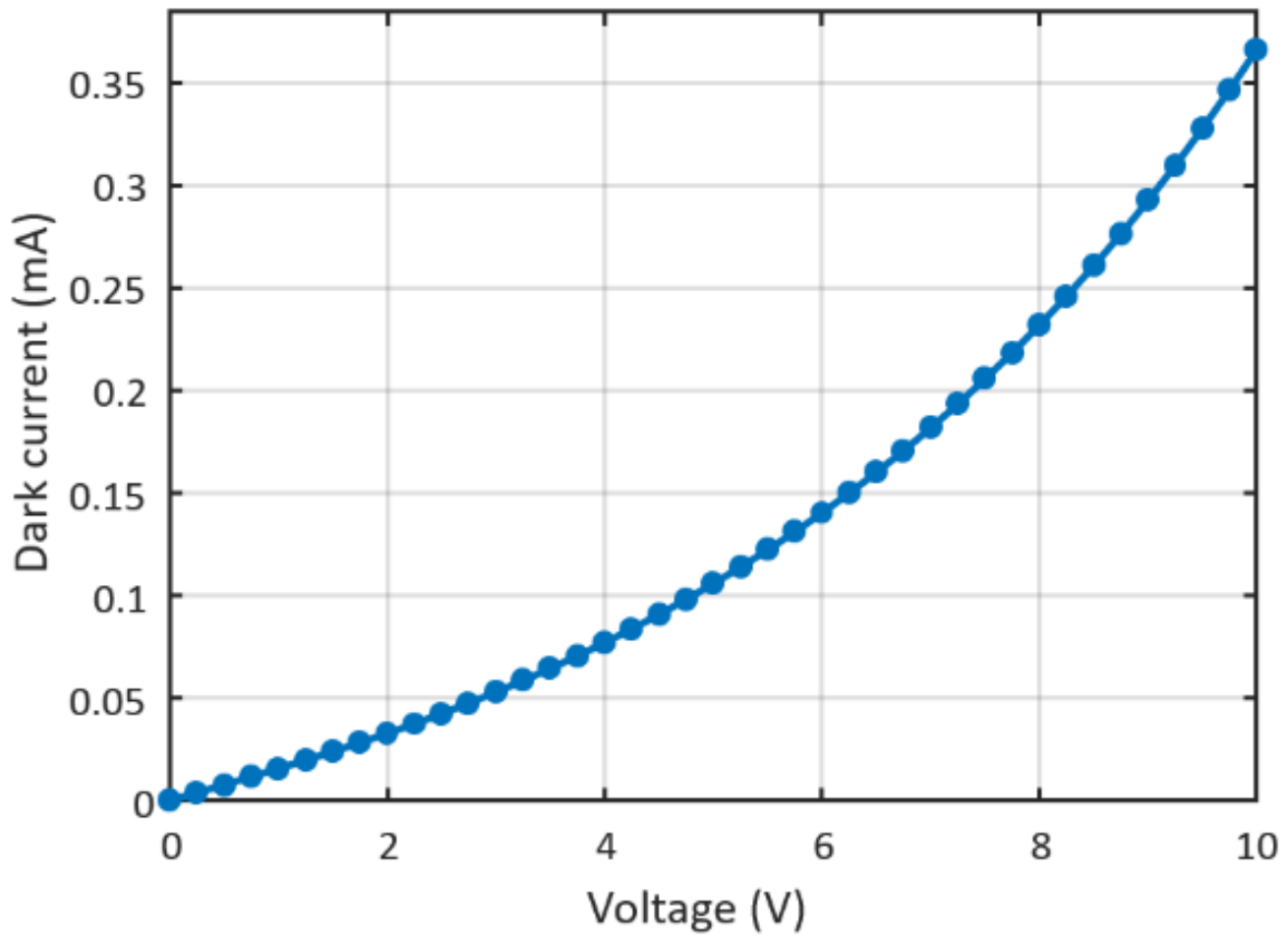

**Figure S4.1 | Dark current of the photomixer under continuous DC bias.**

To reduce the average thermal load and mitigate the risk of thermal breakdown while enabling operation at high instantaneous optical pump power and bias voltage, the optical pump and bias voltage were synchronously modulated at 250 kHz. This pulsed operation provides a thermal relaxation period between successive excitation pulses and enables higher on-state mmWave/THz power levels [S6]. The temporal profile of the modulated optical pump is shown in Fig. S4.2. Due to the finite rise and fall times of the modulation, as well as residual waveform distortion, the measured optical pulse deviates from an ideal rectangular pulse.

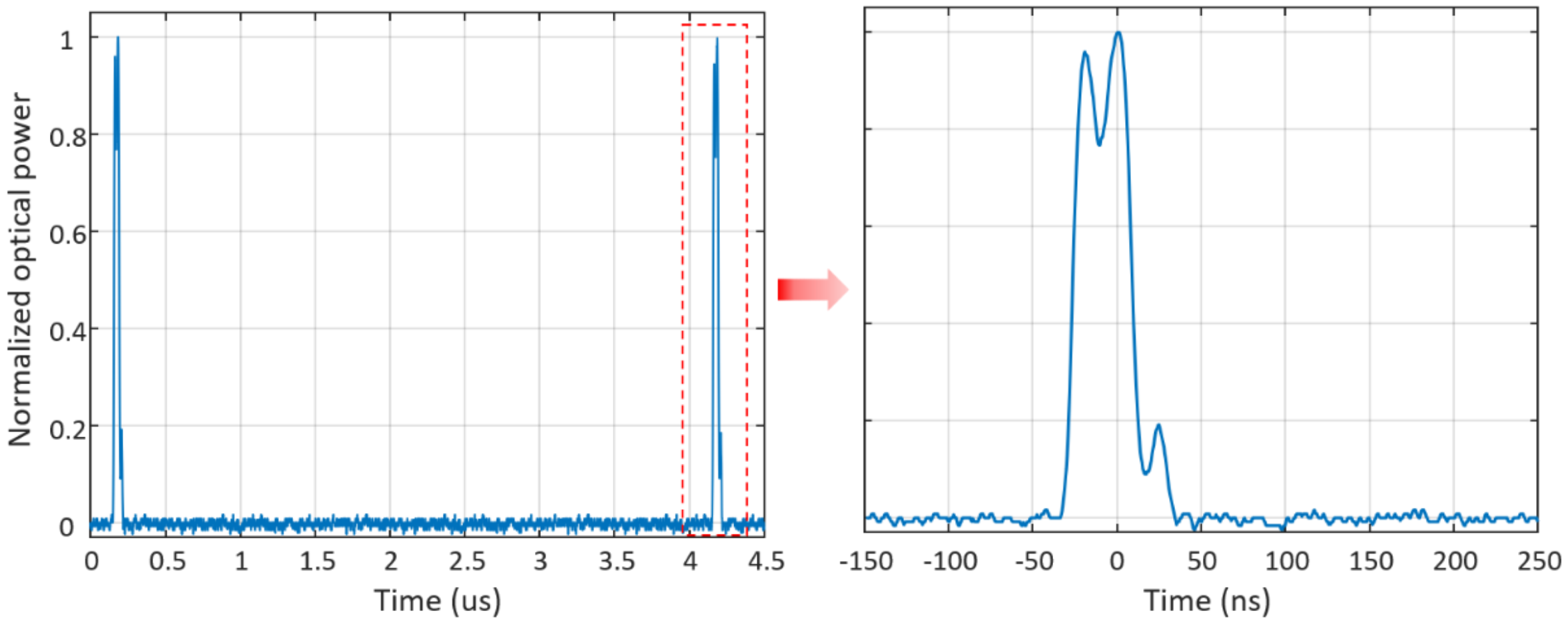

**Figure S4.2 | Temporal profile of the modulated optical pump.** Left: Pulse train measured over one modulation period. Right: Magnified view of a single pulse.

We use the average radiation power measured by a calibrated pyroelectric detector and the radiation pulse duration to determine the mmWave/THz power generated during each radiation cycle. The effective pulse width of the generated radiation is determined from the measured mmWave/THz spectra, as shown in Fig. S4.3. As described in the Methods section, harmonic mixers were used to down-convert the generated radiation to intermediate radio frequencies, and the down-converted spectra were measured using an RF spectrum analyzer. For an ideal rectangular pulse of duration $\tau_{eff}$, the corresponding power spectral envelope follows a $sinc^2$ dependence. By fitting the measured mmWave/THz spectra to this spectral response, an effective rectangular pulse duration of $\tau_{eff} = 40$ ns was obtained. This spectrally inferred effective duration was used to calculate the radiation duty cycle as 40 ns / (1/250 kHz) = 1%. The measured time-averaged power was then converted to the effective mmWave/THz power within each radiation cycle.

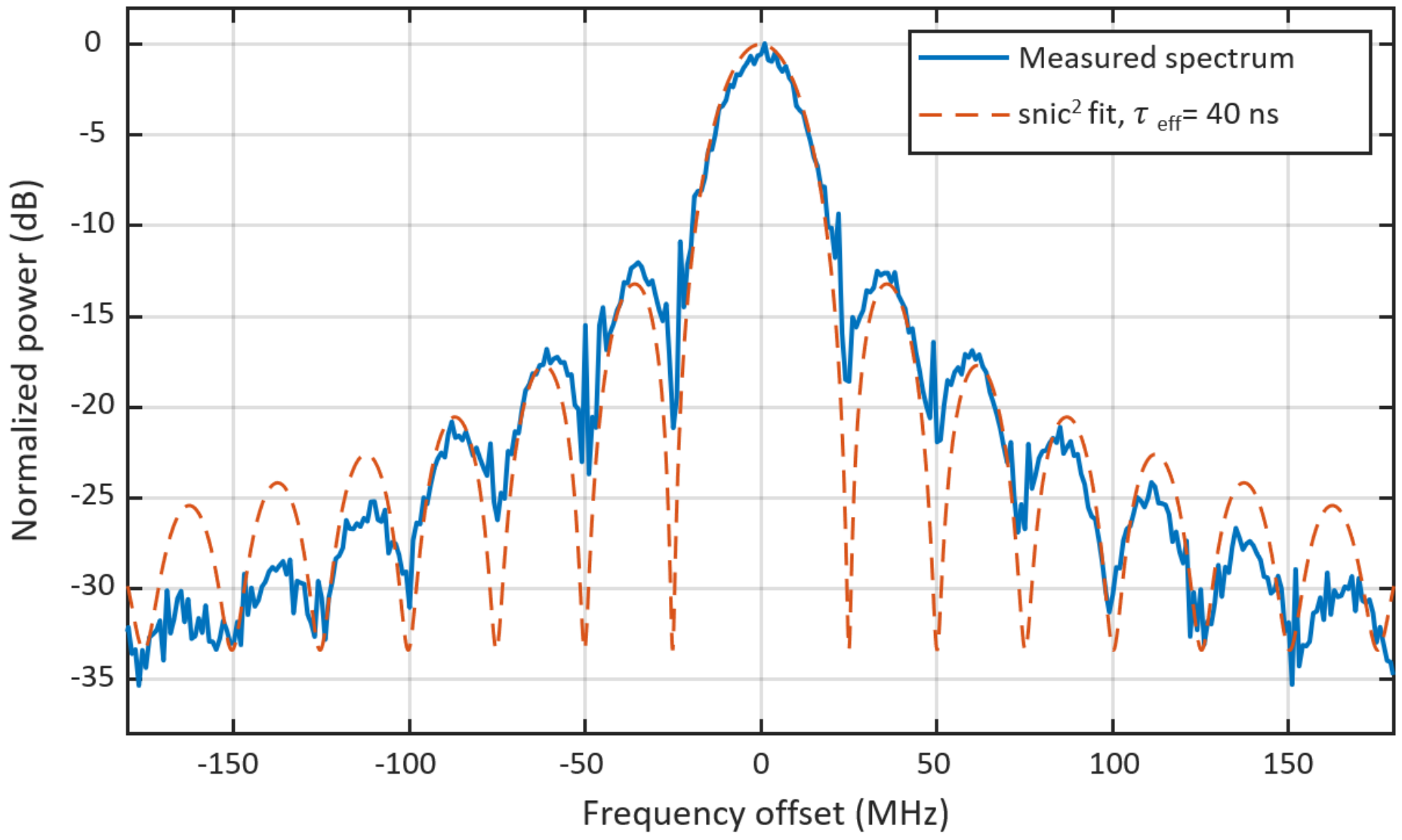

**Figure S4.3 | Normalized mmWave/THz spectral envelope under modulated optical excitation.** The measured spectrum (blue solid curve) is fitted with a $\text{sinc}^2$ spectral response corresponding to an ideal rectangular pulse (red dashed curve), yielding an effective rectangular pulse duration of $\tau_{eff} = 40$ ns.

## Supplementary References

[S1] B. Peng, Ş. K. Özdemir, W. Chen, F. Nori, and L. Yang, What is and what is not electromagnetically induced transparency in whispering-gallery microcavities, Nat. Commun. **5**, 5082 (2014).

[S2] P. Del'Haye, A. Coillet, W. Loh, K. Beha, S. B. Papp, and S. A. Diddams, Phase steps and resonator detuning measurements in microresonator frequency combs, Nat. Commun. **6**, 5668 (2015).

[S3] C. M. Gentry, X. Zeng, and M. A. Popović, Tunable coupled-mode dispersion compensation and its application to on-chip resonant four-wave mixing, Opt. Lett. **39**, 5689 (2014).

[S4] S. A. Miller, Y. Okawachi, S. Ramelow, K. Luke, A. Dutt, A. Farsi, A. L. Gaeta, and M. Lipson, Tunable frequency combs based on dual microring resonators, Opt. Express **23**, 21527 (2015).

[S5] X. Xue, Y. Xuan, P. Wang, Y. Liu, D. E. Leaird, M. Qi, and A. M. Weiner, Normal-dispersion microcombs enabled by controllable mode interactions, Laser Photonics Rev. **9**, (2015).

[S6] C. W. Berry, M. R. Hashemi, S. Preu, H. Lu, A. C. Gossard, and M. Jarrahi, High power terahertz generation using 1550 nm plasmonic photomixers, Appl. Phys. Lett. **105**, 011121 (2014).